\documentclass[a4paper,12pt]{article}
\usepackage{amsfonts,amsthm,amsmath,amssymb,graphicx,hyperref,color,youngtab}

\def\ket#1{\left| #1\right\rangle}

\def\Tr{{\rm Tr\, }}

\newcommand{\be}{\begin{equation}}
\newcommand{\bea}{\begin{eqnarray}}

\newcommand{\ee}{\end{equation}}
\newcommand{\eea}{\end{eqnarray}}

\begin{document}

\makeatletter
\@addtoreset{equation}{section}
\makeatother
\renewcommand{\theequation}{\thesection.\arabic{equation}}

\rightline{}
\vspace{1.8truecm}

\vspace{15pt}


{\LARGE{  
\centerline{\bf LLM Magnons} 
}}  

\vskip.5cm 

\thispagestyle{empty}
    {\large \bf 
\centerline{Robert de Mello Koch\footnote{ {\tt robert@neo.phys.wits.ac.za}}, 
Christopher Mathwin\footnote{ {\tt christopher.mathwin@students.wits.ac.za}}}
\centerline{\large \bf and   Hendrik J.R. van Zyl\footnote{ {\tt hjrvanzyl@gmail.com}}}}

\vspace{.4cm}
\centerline{{\it National Institute for Theoretical Physics ,}}
\centerline{{\it School of Physics and Mandelstam Institute for Theoretical Physics,}}
\centerline{{\it University of Witwatersrand, Wits, 2050, } }
\centerline{{\it South Africa } }

\vspace{1.4truecm}

\thispagestyle{empty}

\centerline{\bf ABSTRACT}

\vskip.4cm 

We consider excitations of LLM geometries described by coloring the LLM plane with concentric black rings.
Certain closed string excitations are localized at the edges of these rings.
The string theory predictions for the energies of magnon excitations of these strings depends on the radii
of the edges of the rings.
In this article we construct the operators dual to these closed string excitations and show how to reproduce
the string theory predictions for magnon energies by computing one loop anomalous dimensions.
These operators are linear combinations of restricted Schur polynomials.
The distinction between what is the background and what is the excitation is accomplished in the choice of 
the subgroup and the representations used to construct the operator.

\setcounter{page}{0}
\setcounter{tocdepth}{2}

\newpage

\tableofcontents

\setcounter{footnote}{0}

\linespread{1.1}
\parskip 4pt

{}~
{}~

\section{Motivation}

There is convincing support in favor of the duality between ${\cal N}=4$ super Yang-Mills theory and IIB strings
on the asymptotically AdS$_5\times$S$^5$ spacetimes\cite{Maldacena:1997re}.
The duality identifies a gauge theory operator for each state in the string theory, and the dimension of this operator with the
energy of the string theory state.
By comparing operator dimensions and string state energies, we find a wealth of non-trivial predictions that can be tested.
Since the duality is a strong/weak coupling relation, the results of perturbative field theory and perturbative string theory are 
not related by the duality and checking the equality of dimensions and energies is in general highly nontrivial.
In the case that we consider, gauge theory operators belonging to the $SU(2|3)$ sector of the theory, this check can be 
carried out in complete detail.
This remarkable progress is possible by exploiting the  symmetry of the problem in an 
interesting way\cite{Beisert:2005tm,Beisert:2006qh}.

The gauge theory operator corresponding to a closed string state is a single trace operator\cite{Berenstein:2002jq} 
constructed using a very large number ($J\sim O(\sqrt{N})$)) of scalar $Z$ fields, with a few impurities (which may be the scalars
$X$, $Y$, $X^\dagger$ or $Y^\dagger$ or of one of four possible fermions).
The symmetry which plays a central role is an $SU(2|2)^2$ symmetry which acts on the impurities.
When the impurities are well separated within the trace, each impurity transforms as a short multiplet of a centrally extended 
$SU(2|2)^2$ symmetry; the fact that the multiplet is short determines its anomalous dimension\cite{Beisert:2005tm}.
The anomalous dimension of the single trace operator is then given by summing the dimensions of the impurities.
The sum of the new magnon central charges\footnote{i.e. the charges switched on for the representations of the magnon but
not present in the original group.} vanishes so that the closed string state is a representation of the original algebra.
An inspired explanation for these dimensions was given in \cite{Berenstein:2005jq}.
This physically motivated description matches the string description, which was worked out in \cite{Hofman:2006xt}.
Exactly the same symmetry algebra plays a role and the central charges of the algebra acquire an elegant geometrical
interpretation as we now explain.
In general, the anticommutator of two supersymmetries in 10 dimensional supergravity includes a gauge transformation 
of the NS-$B_{\mu\nu}$ field, which acts non-trivially on stretched strings.
The parameter of the gauge transformation was computed in \cite{Schwarz:1983qr} in terms of the Killing spinors of the geometry.
By inserting the explicit expression of the LLM Killing spinors\cite{Lin:2004nb} into the general formula for the gauge 
transformation produced by the anticommutator of two supercharges, one finds that the relevant NS gauge transformations on 
the LLM plane, are those with a constant gauge parameter. 
Thus any string stretched along the LLM plane will acquire a phase under these gauge transformations.
These are the central charges that we are after\cite{Hofman:2006xt}.
To develop the geometrical interpretation of these central charges, we need to develop the geometrical picture of the string
worldsheet projected to the LLM plane.
The $Z$s inside the trace carry the quantum numbers of gravitons, which map to a single point, orbiting in a circle of radius $r=1$.
The impurities inside the trace map into ``giant magnons'' - straight line segments that stretch between the points on the
$r=1$ circle where the $Z$s are located\cite{Hofman:2006xt}. 
Thus, magnons are represented by directed line segments on the plane.
Introduce a complex number $k$ for each oriented segment, whose magnitude is the length of this segment and whose phase is 
the direction of this segment.
Thanks to the fact that the gauge transformations giving rise to the central charges are constant, the central charge of the
corresponding magnon is $k$\cite{Hofman:2006xt}.
The projected closed string is a polygon on the LLM plane with a consistent assignment of orientation to each side of the polygon,
so that the central charges of the magnons sum to zero.

The powerful $SU(2|2)^2$ symmetry arguments developed above do not make use of the planar approximation.
It is then natural to expect that the argument continues to work even for operators with such a large dimension that
the planar approximation is no longer accurate\footnote{In this case there is in general no integrability and so the constraints
implied by the $SU(2|2)^2$ analysis give results that would be difficult to obtain by any other method.}.
A natural setting in which this expectation can be tested, is for open strings attached to giant gravitons.
In this case, one is studying correlators of operators that have a bare dimension of order $N$.
This has been carried out for the maximal giant in \cite{Hofman:2007xp} and in complete generality (any number of giants and 
dual giants of any size) in \cite{Koch:2015pga}.
These studies confirm that the $SU(2|2)^2$ symmetry arguments continue to work.
Another natural situation to consider is that of closed strings (and/or giant gravitons) exploring the LLM geometries.
These correspond to operators with a bare dimension of order $N^2$.
In this article we study this general class of problems.

The LLM geometries are regular ${1\over 2}$ BPS solutions of type IIB string theory that are asymptotically AdS$_5\times$S$^5$.
They are dual to operators constructed using a single complex field $Z$.
These geometries enjoy an $R\times SO(4)\times SO(4)$ isometry group and have a metric which is given by ($i,j=1,2$)
\bea
   ds^2= -h^{-2}(dt+V_i dx^i)^2 +h^2 (dy^2 +dx^idx^i)+ye^G d\Omega_3^2 + ye^{-G}d\tilde{\Omega}_3^2 \, ,
\eea
where
\bea
    h^{-2}=2y\cosh G,\qquad z={1\over 2}\tanh G,
\eea
\bea
    y\partial_y V_i=\epsilon_{ij}\partial_j z,\qquad y(\partial_i V_j-\partial_j V_i)=\epsilon_{ij}\partial_y z .
\eea
Thus, the metric is completely determined by the function $z$, which is a function of $y,x^1$ and $x^2$.
It is obtained by solving
\bea
   \partial_i\partial_i z+y\partial_y {\partial_y z\over y}=0 .
\eea
Regularity requires that $z=\pm {1\over 2}$ on the LLM ($y=0$) plane. 
As a consequence, the complete set of LLM solutions can be labeled by coloring the entire LLM plane into black 
(where $z=-{1\over 2}$) and white (where $z=+{1\over 2}$) regions\cite{Lin:2004nb}. 

Each LLM geometry corresponds to a specific ${1\over 2}$ BPS operator. 
We will restrict ourselves to geometries that correspond to coloring the plane with a collection of concentric annuli, surrounding
a central disk which may be of either color.
Each such geometry corresponds to a Schur polynomial, labeled by a Young diagram $B$\cite{Corley:2001zk,Berenstein:2004kk}.
Every plane coloring we consider can be translated into a Young diagram and hence into a definite gauge theory operator.
An example of the translation is shown in Fig \ref{fig:rings}.
The AdS$_5\times$S$^5$ geometry corresponds to a black disk of radius 1.
Gravitons dual to gauge theory operators built using only the $Z$ field would follow circular orbits at the outer edge
of any black region.
We would also have closed string states, with worldsheet given by a polygon with every vertex on the outer edge of any
black annulus. 
Each polygon edge is a magnon.
Using the general form of the LLM Killing spinors, it is again true that the anticommutator of two supersymmetries includes
a constant gauge transformation.
Consequently the central charge of the magnon is still set by the corresponding line segment.
The angle that the line segment corresponding to a magnon subtends with respect to the origin of the LLM plane,
determines the momentum of the magnon.
To turn this angle into a length (and hence an energy) we need to know the radius of the circle(s) that the line segment's
end points are located on.
Thus, the values of the central charges as well as the dispersion relation depend on the radii of the annuli in the LLM boundary
condition.
\begin{figure}[ht]%
\begin{center}
\includegraphics[width=0.9\columnwidth]{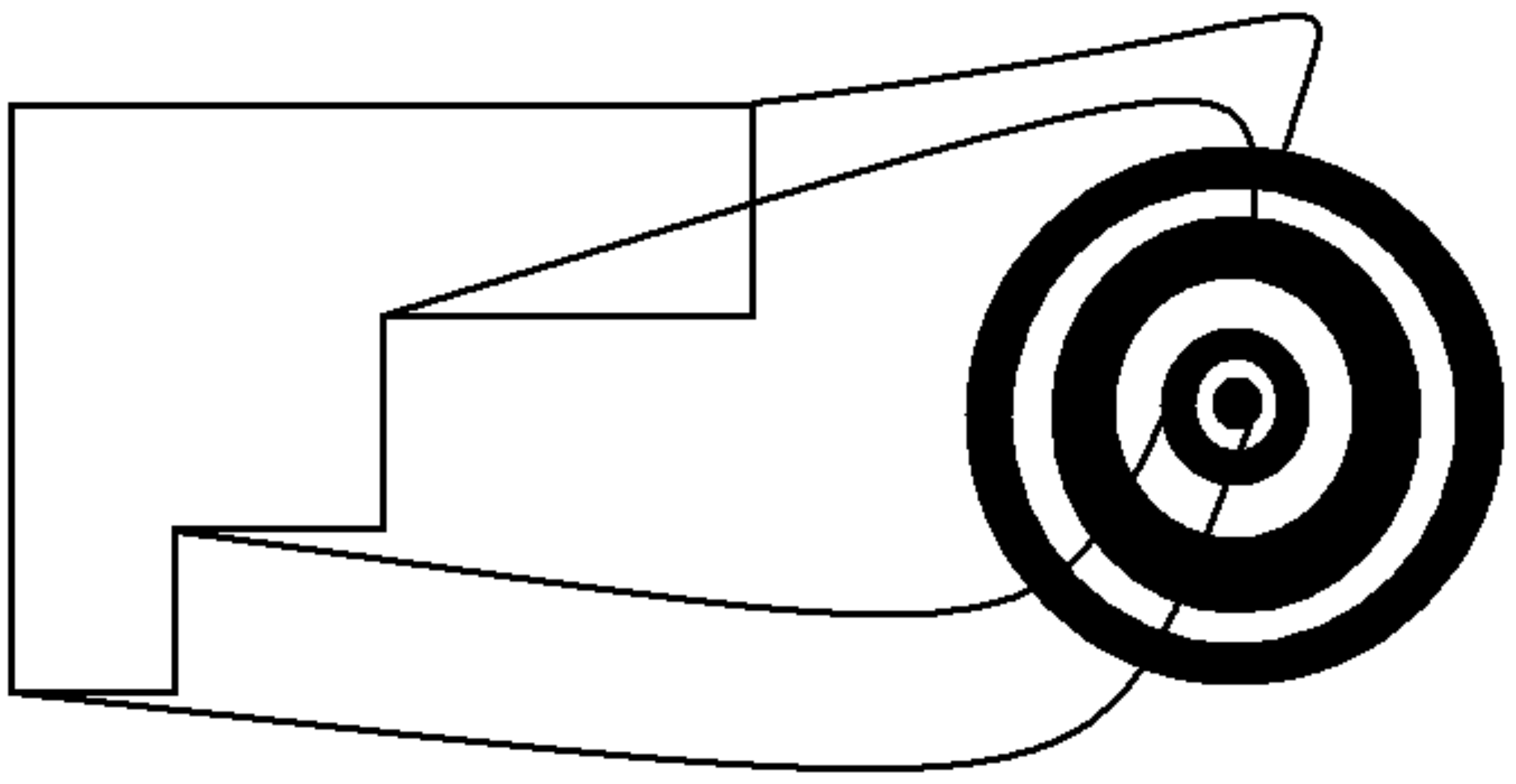}%
\caption{Inward pointing corners correspond to outer radii of black regions.
Each black (white) region corresponds to a vertical (horizontal) line on the edge of the Young diagram.
The area of each white (black) region divided by $\pi$ is given by the number of columns (rows)
in the corresponding line, divided by $N$.
The area of the central black disk divided by $\pi$ is given by $N$ minus the total number of rows, divided by $N$.
The total area of the black regions is $\pi$.}%
\label{fig:rings}%
\end{center}
\end{figure}
 
Our goal is to write down the operators dual to closed strings in the above LLM geometries. 
The first question we need to address is how to write down gauge theory operators that are localized at the edge of an annulus
in the dual gravity description.
Once these operators are constructed, we can consider the problem of determining their anomalous dimensions.
By computing their one loop anomalous dimensions, we can determine the central charges of the representations in which the
magnons transform and compare to the string theory predictions.
We will find complete agreement.
This both gives support that we have correctly constructed localized closed string states and that the $SU(2|2)^2$ symmetry
analysis is still applicable in this case, as one would expect.
For some of the closed string operators we consider, the problem of determining the anomalous dimensions is an integrable
problem and the results obtained in the planar limit generalize immediately without any effort.

This article is organized as follows: In section 2 we study the planar operator mixing problem at one loop.
This discussion is usually phrased in terms of single trace operators, a language which is not useful outside of the planar
limit.
We translate this discussion into the language of restricted Schur polynomials.
It is the restricted Schur language that will generalize.
In section 3 we point out, using simple examples, some of the issues related to constructing operators dual to
excitations localized on the LLM plane.
Using these insights we give our proposal for operators dual to localized closed string states in section 4 and we compute their 
anomalous dimensions in section 5.
These all correspond to excitations localized at the outer edge of an LLM annulus.
Although our computations are quite technical and make heavy use of group representation theory the result is striking in
its simplicity: the problem in the nontrivial geometry is given by simply scaling the $N$ dependence in the planar result.
In section 6 we consider an example that is simple enough that it can be treated without any use of group representation theory
and we confirm our results in this simple setting.
In section 7 we explain how to describe excitations localized at the inner edge of an LLM annulus.
Finally, in section 8 we discuss our results and draw some conclusions.

\section{Dilatation Operator in the Planar Limit using Restricted Schur Polynomials}

To simplify our discussion we will restrict ourselves to the $su(2)$ sector of ${\cal N}=4$ super Yang-Mills theory.
Restricting to this sector does not interfere with our goal of computing the central charge appearing in the $SU(2|2)$ symmetry
algebra, but it will significantly simplify our arguments.
The local operators of interest to us are loops that correspond to closed string states\cite{Berenstein:2002jq}.
In the planar limit, a basis for these operators is labeled by the ordered set of integers $\{n_k\}$
\bea
   O(\{ n_k\})=  {\rm Tr}(Z^{n_1}YZ^{n_2}Y\cdots Z^{n_m}Y)={\rm Tr}(\sigma Z^{\otimes n}Y^{\otimes m})
  \label{loop}
\eea
with $\sum_k n_k=n$.
Not all sets correspond to distinct operators due to cyclicity of the trace.
For string states we would hold $n+m\sim \sqrt{N}$ and for the states we are interested in, we take $m\ll n$.
The planar approximation is valid as long as ${(m+n)^2\over N}\ll 1$.
We refer to these as ``loop operators'' and make use of the following notation
\bea
{\rm Tr} (\sigma Z^{\otimes n}Y^{\otimes m})=
Y^{i_1}_{i_{\sigma(1)}}\cdots Y^{i_m}_{i_{\sigma(m)}}Z^{i_{m+1}}_{i_{\sigma(m+1)}}\cdots Z^{i_{n+m}}_{i_{\sigma(n+m)}}\, .
\eea
Each of the operators (\ref{loop}) corresponds to a permutation that is a single $m+n$ cycle
\bea
   O(\{ n_k\})=  {\rm Tr} (\sigma_{\{ n_k\}} Z^{\otimes n}Y^{\otimes m})
\eea
The loop operators are a particularly convenient description of the planar limit.
In particular, it was in terms of these variables that the link to the spin chain and the subsequent discovery of integrability
was made\cite{Minahan:2002ve,Beisert:2010jr}.
This is a consequence of the fact that there is a bijection between loop operators and spin chain states.
In addition, these variables also provide an explicit and direct link to the string worldsheet\cite{Kruczenski:2003gt}.

In the planar limit the loop operators are orthogonal.
As we increase $n+m$ beyond $O(\sqrt{N})$, the loop operators start to mix and they no longer provide a useful 
description\cite{Balasubramanian:2001nh}.
The operators we are interested in correspond to closed string states in an LLM geometry\cite{Lin:2004nb}, 
so that $n\sim O(N^2)$ and ${m\over n}\ll 1$.
The loop operators are useless in this limit.
The basis provided by the restricted Schur polynomials is far more useful: these operators are exactly orthogonal in the free field
theory and their mixing at loop level is tightly constrained\cite{Bhattacharyya:2008rb,Koch:2011hb}\footnote{There are
a number of distinct bases that are exactly orthogonal in the free field
 limit\cite{Brown:2007xh,Brown:2008ij,Kimura:2007wy,Kimura:2009jf,Kimura:2012hp,Pasukonis:2013ts}.}.
The definition of the restricted Schur polynomial is\cite{Bhattacharyya:2008rb}
\begin{equation}
\chi_{R,(r,s)\alpha\beta}(Z,Y)={1\over n!m!}\sum_{\sigma\in S_{n+m}}\chi_{R,(r,s),\alpha\beta}(\sigma )
Y^{i_1}_{i_{\sigma (1)}}\cdots Y^{i_m}_{i_{\sigma (m)}}
Z^{i_{m+1}}_{i_{\sigma (m+1)}}\cdots Z^{i_{n+m}}_{i_{\sigma (n+m)}}\, .
\label{restrictedschur}
\end{equation}
In this definition $R$ is a Young diagram with $n+m$ boxes and hence labels an irreducible representation (irrep) of $S_{n+m}$,
$r$ is a Young diagram with $n$ boxes and labels an irrep of $S_{n}$ and
$s$ is a Young diagram with $m$ boxes and labels an irrep of $S_{m}$. 
The group $S_{n+m}$ has an $S_n\times S_m$ subgroup. 
Taken together $r$ and $s$ label an irrep of this subgroup.
A single irrep $R$ will in general subduce many possible representations of the subgroup. 
A particular irrep of the subgroup may be subduced more than once in which case we
must introduce a multiplicity label to keep track of the different copies subduced. 
The indices $\alpha$ and $\beta$ appearing above are these multiplicity labels. 
The object $\chi_{R,(r,s)\alpha\beta}(\sigma )$ is called a restricted character\cite{de Mello Koch:2007uu}. 
At present there are no general powerful methods to compute these characters and this is one of the main obstacles that
must be overcome when performing explicit computations.
In this section there are a number of useful identities that we will prove, obtained by writing the known planar action of the 
dilatation operator on loop operators, in terms of the restricted Schur polynomial operators.
In the Appendices we will derive these results, and more general exact identities, using the representation theory
of the symmetric group.
These identities are all we will need in this paper, so that we entirely avoid the need to explicitly evaluate any 
restricted characters.

The restricted Schur polynomials provide a complete basis.
As a consequence we can write the loop operators as a linear combination of restricted Schur polynomials.
The basic identity we need to carry this out is\cite{Bhattacharyya:2008xy}
\bea
{\rm Tr} (\sigma Z^{\otimes n}Y^{\otimes m})=\sum_{T,(t,u)\alpha\beta}
{d_T n! m!\over d_t d_u (n+m)!}\chi_{T,(t,u)\alpha\beta}(\sigma^{-1})
\chi_{T,(t,u)\beta\alpha}(Z,Y)
\eea
Applying this to a loop operator, we find
\bea
O(\{ n_k\})&=&\sum_{T,(t,u)\alpha\beta}
{d_T n! m!\over d_t d_u (n+m)!}\chi_{T,(t,u)\alpha\beta}(\sigma^{-1}_{\{n_k\}})
\chi_{T,(t,u)\beta\alpha}(Z,Y)\cr
&=&\sum_{T,(t,u)\alpha\beta}
\sqrt{f_T{\rm hooks}_t{\rm hooks}_u\over {\rm hooks}_T}\chi_{T,(t,u)\alpha\beta}(\sigma^{-1}_{\{n_k\}})
O_{T,(t,u)\beta\alpha}(Z,Y)
\label{YoungLoop}
\eea
where $O_{T,(t,u)\beta\alpha}(Z,Y)$ is a restricted Schur polynomial normalized to have unit two point function.

The action of the dilatation operator on the loop operators in the planar limit is remarkably simple.
The one loop dilatation operator, in the $SU(2)$ sector, is\cite{charlotte}
\bea
   D=-{g_{YM}^2\over 8\pi^2}{\rm Tr}\left(\left[ Y,Z\right]\left[{d\over dY},{d\over dZ}\right]\right)
\eea
Acting on the loop operators we have
\bea
   DO(\{ n_k\})
={g_{YM}^2 N\over 8\pi^2}
\left(2O(\{n_k\})-O(\{n_1+1,n_2-1,\cdots,n_m\})-O(\{n_1-1,n_2+1,\cdots,n_m\})
\right)\cr
+{g_{YM}^2 N\over 8\pi^2}
\left(2O(\{n_k\})-O(\{n_1,n_2+1,n_3-1,\cdots,n_m\})-O(\{n_1,n_2-1,n_3+1,\cdots,n_m\})
\right)\cr
+ \qquad\qquad\cdots\cdots\cdots
\cr
+{g_{YM}^2 N\over 8\pi^2}
\left(2O(\{n_k\})-O(\{n_1-1,n_2,\cdots,n_m+1\})-O(\{n_1+1,n_2,\cdots,n_m-1\})
\right)\cr
\label{actonloop}
\eea
This result is certainly not exact - only the planar contractions are retained.
The subleading (non-planar) terms will induce a splitting of the above loop operator into a double trace operator.

When acting on restricted Schur polynomials the one loop dilatation operator takes the form\cite{DeComarmond:2010ie}
\bea
  DO_{R,(r,s)\mu_1 \mu_2}(Z,Y)=\sum_{T,(t,u)\nu_1\nu_2} N_{R,(r,s)\mu_1 \mu_2;T,(t,u)\nu_1\nu_2}O_{T,(t,u)\nu_1\nu_2}(Z,Y)
\eea
where
\begin{eqnarray}
\label{dilat}
N_{R,(r,s)\mu_1 \mu_2;T,(t,u)\nu_1\nu_2}&&\!\!\!\!\!\!\!\!= - {g_{YM}^2\over 8\pi^2}
\sum_{R'}{c_{RR'} d_T n m\over d_{R'} d_t d_u (n+m)}
\sqrt{f_T \, {\rm hooks}_T\, {\rm hooks}_r \, {\rm hooks}_s \over f_R \, {\rm hooks}_R\, {\rm hooks}_t\, {\rm hooks}_u}\times
\\
\nonumber
&&\!\!\!\!\!\!\!\!\!\!\!\!\!\!\!\!\!\!\!\!\!\!\!\!\!\!\!\!\!\!\!\!\!\!\!\!\!\!\!\!\!\!
\times{\rm Tr}\Big(\Big[ \Gamma^{(R)}((1,m+1)),P_{R,(r,s)\mu_1 \mu_2}\Big]I_{R'\, T'}
\Big[\Gamma^{(T)}((1,m+1)),P_{T,(t,u)\nu_2\nu_1}\Big]I_{T'\, R'}\Big) 
\end{eqnarray}
The trace above is over irrep $R\vdash n+m$. $I_{R'\, T'}$ is a map from irrep $R'$ to irrep $T'$, where both are irreps
of $S_{n+m-1}$. 
$R'$ is one of the irreps subduced from $R$ upon restricting 
to the $S_{n+m-1}$ subgroup of $S_{n+m}$ obtained by keeping
only permutations that obey $\sigma (1)=1$. $T'$ is subduced by $T$ in the same way.
$I_{R'\, T'}$ is only non-zero if $R'$ and $T'$ have the same shape.
Thus, to get a non-zero result $R$ and $T$ must differ at most by the placement of a single box.
$d_a$ denotes the dimension of symmetric group irrep $a$.
$f_S$ is the product of the factors in Young diagram $S$ and 
${\rm hooks}_S$ is the product of the hook lengths of Young diagram $S$.
Finally, $c_{RR'}$ is the factor of the corner box that must be removed from $R$ to obtain $R'$.
Combining the above results, we now find
\bea
  DO(\{ n_k\})&=&\sum_{R,(r,s)\mu_1 \mu_2}
\sqrt{f_R{\rm hooks}_r{\rm hooks}_s\over {\rm hooks}_R}
\chi_{R,(r,s)\mu_2\mu_1}(\sigma^{-1}_{\{n_k\}})
D O_{R,(r,s)\mu_1\mu_2}(Z,Y)\cr
&&\!\!\!\!\!\!\!\!\!\!\!\!\!\!\!\!\!\!\!\!\!\!\!\!\!\!\!\!\!\!\!\!
=\sum_{T,(t,u)\nu_1\nu_2}\sum_{R,(r,s)\mu_1 \mu_2}
\sqrt{f_R{\rm hooks}_s\over {\rm hooks}_{R/r}}\chi_{R,(r,s)\mu_2\mu_1}(\sigma^{-1}_{\{n_k\}})
N_{R,(r,s)\mu_1 \mu_2;T,(t,u)\nu_1\nu_2} O_{T,(t,u)\nu_1\nu_2}(Z,Y)\cr
&&\label{actonschur}
\eea
where
\bea
    {\rm hooks}_{R/r}={{\rm hooks}_{R}\over {\rm hooks}_{r}}
\eea
The equality of (\ref{actonloop}) and (\ref{actonschur}) now proves that
\bea
&&\sum_{T,(t,u)\nu_1\nu_2}\sum_{R,(r,s)\mu_1 \mu_2}
\sqrt{f_T{\rm hooks}_u\over {\rm hooks}_{T/t}}\chi_{T,(t,u)\nu_2\nu_1}(\sigma^{-1}_{\{n_k\}})
N_{T,(t,u)\nu_1 \nu_2;R,(r,s)\mu_1\mu_2} O_{R,(r,s)\mu_1\mu_2}(Z,Y)\cr
&&={g_{YM}^2 N\over 8\pi^2}
\sum_{R,(r,s)\mu_1 \mu_2}
\sqrt{f_R{\rm hooks}_s\over {\rm hooks}_{R/r}}O_{R,(r,s)\mu_1\mu_2}(Z,Y)\cr
&&\quad \Big(
\chi_{R,(r,s)\mu_2\mu_1}(2\sigma^{-1}_{\{n_k\}}-\sigma^{-1}_{\{n_1+1,n_2-1,\cdots,n_m\}}
-\sigma^{-1}_{\{n_1-1,n_2+1,\cdots,n_m\}})
\cr
&&\qquad+\chi_{R,(r,s)\mu_2\mu_1}(2\sigma^{-1}_{\{n_k\}}-\sigma^{-1}_{\{n_1,n_2+1,n_3-1,\cdots,n_m\}}
-\sigma^{-1}_{\{n_1,n_2-1,n_3+1,\cdots,n_m\}})+\cdots+\cr
&&\qquad+\chi_{R,(r,s)\mu_2\mu_1}(2\sigma^{-1}_{\{n_k\}}-\sigma^{-1}_{\{n_1-1,n_2,\cdots,n_m+1\}}
-\sigma^{-1}_{\{n_1+1,n_2,\cdots,n_m-1\}})\Big)\cr
&&
\eea
Finally, since the restricted Schur polynomials are independent, this implies
\bea
&&\sum_{T,(t,u)\nu_1\nu_2}
\sqrt{f_T{\rm hooks}_u\over {\rm hooks}_{T/t}}\chi_{T,(t,u)\nu_2\nu_1}(\sigma^{-1}_{\{n_k\}})
N_{T,(t,u)\nu_1 \nu_2;R,(r,s)\mu_1\mu_2}\cr
&&={g_{YM}^2 N\over 8\pi^2}
\sqrt{f_R{\rm hooks}_s\over {\rm hooks}_{R/r}}\Big(
\chi_{R,(r,s)\mu_2\mu_1}(2\sigma^{-1}_{\{n_k\}}-\sigma^{-1}_{\{n_1+1,n_2-1,\cdots,n_m\}}
-\sigma^{-1}_{\{n_1-1,n_2+1,\cdots,n_m\}})
\cr
&&\qquad+\chi_{R,(r,s)\mu_2\mu_1}(2\sigma^{-1}_{\{n_k\}}-\sigma^{-1}_{\{n_1,n_2+1,n_3-1,\cdots,n_m\}}
-\sigma^{-1}_{\{n_1,n_2-1,n_3+1,\cdots,n_m\}})+\cdots+\cr
&&\qquad+\chi_{R,(r,s)\mu_2\mu_1}(2\sigma^{-1}_{\{n_k\}}-\sigma^{-1}_{\{n_1-1,n_2,\cdots,n_m+1\}}
-\sigma^{-1}_{\{n_1+1,n_2,\cdots,n_m-1\}})\Big)\cr
&&\label{restrictedCharId}
\eea
which is the identity obeyed by restricted characters that we aimed to derive.
We stress that this is not an exact result - we have used the simplifications of the planar limit to obtain it.

This rewriting of the planar dilatation operator is interesting. 
The above identity is written using Young diagrams and the language of restricted characters.
However, the appearance of the permutations which label the loop operators keeps manifest the bijection to the spin chain.
This is the restricted Schur polynomial way of mapping to the spin chain dynamics.
In what follows, we will argue that the description of closed string states using a permutation is a useful description for
operators dual to closed strings probing LLM backgrounds.
The operators described using the permutation are certainly not single trace operators.
Indeed, in the next section we will argue that single trace operators in the gauge theory are not dual to localized excitations
in the dual gravity.
This discussion will motivate the form of the operators that are dual to closed strings in the LLM geometries.

\section{How not to Localize}

The operators dual to the closed string states we study are composed of collections of $Z$s raised to some power, separated 
by $Y$ fields. 
The worldsheet geometry of these strings has been studied in \cite{Hofman:2006xt}. 
The pieces of the worldsheet constructed by the $Z$s are localized on a particular circle in the LLM plane, 
while the $Y$ fields correspond to worldsheet magnons that stretch between these points. 
In this section we consider the problem of writing operators, composed entirely out of $Z$s, that are localized in the radial
direction of the LLM plane.
Using the intuition gathered from this toy problem we will construct operators dual to closed strings in the LLM geometries.
This section is largely a review of relevant results from \cite{Koch:2008ah,Koch:2008cm,deMelloKoch:2009jc}.
The papers \cite{Berenstein:2005aa,Vazquez:2006id,Chen:2007gh,Lin:2010sba} include closely related ideas.
These articles used an eigenvalue density description for the ${1\over 2}$ BPS operators in the gauge theory.
The construction of localized giant graviton branes has been pursued in
\cite{Berenstein:2013md,Berenstein:2013eya,Berenstein:2014pma,Berenstein:2014isa,Berenstein:2014zxa}
using a collective coordinate description.

Consider the AdS$_5\times$S$^5$ background, which is dual to the vacuum state of ${\cal N}=4$ super Yang-Mills theory.
Point like gravitons, with a momentum $p\sim O(1)$ are dual to operators
\bea
   {\rm Tr}(Z^p)
\eea
They are localized at the radius $r=1$ on the LLM plane.
Thus, in this case, a single trace operator with $O(1)$ fields is dual to an object in the string theory, localized in the radial direction.

Single trace operators do not create localized states in general\cite{Koch:2008ah}. 
Consider a three ring geometry - two concentric rings with a central black disk. 
This corresponds to a Schur polynomial labeled by a Young diagram with the following shape
\bea
R={\tiny \yng(12,12,12,12,12,12,6,6,6,6,6,6)}
\eea
Applying ${\rm Tr}(Z)=\chi_{\tiny\yng(1)}(Z)$ to the above state, the product $\chi_R(Z)\chi_{\tiny\yng(1)}(Z)$ is easily
computed using the Littlewood Richardson rule\cite{Corley:2001zk}.
The product consists of three terms, one labeled by a Young diagram obtained by adding a box to row 1 of $R$,
one labeled by a Young diagram obtained by adding a box to row 7 of $R$ and one labeled by a Young diagram obtained
by adding a box to row 13 of $R$.
In the dual gravity this is a superposition of states. One state is the original geometry with a single graviton localized at the 
outer edge of the largest ring, one state is the original geometry with a single graviton localized at the outer edge of the smallest
ring and the third state is the original geometry with a single graviton localized at the outer edge of the central 
disk.  
Thus, the single trace state can not be interpreted as an excitation of the original geometry, localized at some radius in the plane.
This completely local and gauge invariant operator in the gauge theory maps to something non-local in the gravity dual.

To localize the excitation in the geometry, we need to mix the indices of the operator creating the excitations with the
indices of the fields making up the background in such a way that the boxes describing the excitation are only added at
one location on the Young diagram describing the original geometry\cite{Koch:2008ah,Koch:2008cm,deMelloKoch:2009jc}.
We need to mix the gauge group indices of the excitation and the background to produce a local excitation.
The details of how these indices are mixed determines where the excitation is localized.
We will not pursue the problem of explicitly constructing operators which create localized excitations further.

\section{Localized Closed String States}\label{LocalizedString}

Using the lessons of the previous section, we will now write down operators dual to closed string states localized on the LLM plane.
We can write the loop operator as a linear combination of restricted Schur polynomials labeled by a triple of Young diagrams 
using the general result (\ref{YoungLoop}) which is true for any permutation $\sigma$.
Our strategy is to write a local version of the restricted Schur polynomial, and then use (\ref{YoungLoop}) to obtain local loops.
We will now motivate and explain our proposal for the localized version of a restricted Schur polynomial in the LLM backgrounds.
Our arguments in what follows holds for general values of $m$ and $n_C$.
For simplicity we will however often consider an example with $m=3$ and $n_C=3$ before stating the general result.

Consider a restricted Schur polynomial with labels
\bea
   R={\tiny \yng(3,2,1)}\qquad r={\tiny \yng(2,1)}\qquad s={\tiny \yng(2,1)}
\eea
These would be summed (along with other possible restricted Schur polynomials) to produce a loop operator 
with three $Y$s and three $Z$s.
To completely specify the possible operators we need a multiplicity label since two copies of ${\tiny \yng(2,1),\yng(2,1)}$ arise 
when we restrict ${\tiny\yng(3,2,1)}$ to the $S_3\times S_3$ subgroup.
We do the restriction by removing boxes from $R$ to leave $r$.
The removed boxes are then assembled to produce $s$.
This way of embedding the representations that appear is particularly convenient, since it allows us to associate the multiplicity label
with $s$ matching the construction developed in \cite{Koch:2011hb}. 
Thus, the possible restricted Schur polynomials are
\bea
   \chi_{{\tiny \yng(3,2,1) (\yng(2,1),\yng(2,1))}\alpha\beta}\qquad \alpha,\beta=1,2
  \label{possSchurs}
\eea
As a consequence of the fact that two copies of ${\tiny \yng(2,1),\yng(2,1)}$ arise, there are four possible operators 
we can define.
We want to write down local versions of these restricted Schur polynomials, in the background corresponding to Young
diagram $B$ given below.
$B$ corresponds to two concentric black rings surrounding a black disk.
We will write down local verions of the above four operators localized around the outer edge of the middle ring. 
These operators are again given by restricted Schur polynomials. $R$ is replaced by the Young diagram obtained by
adjoining $R$ to $B$ in the appropriate location and similarly for $r$. 
The new labels are denoted $R_B$ and $r_B$. 
$s$ is unchanged and since the problem of assembling the boxes removed from $R_B$ to form $s$ has exactly the same
multiplicity as the problem of assembling the boxes removed from $R$ to form $s$ we assign it a multiplicity label that takes
the same values as the original label did.
For the example we are considering we have
\bea
B={\tiny \yng(14,14,14,14,14,14,14,5,5,5,5,5)}\qquad R={\tiny \yng(3,2,1)}\qquad r={\tiny \yng(2,1)}
\eea
and
\bea
R_B={\tiny \yng(14,14,14,14,14,14,14,8,7,6,5,5)}
\qquad
r_B={\tiny \yng(14,14,14,14,14,14,14,7,6,5,5,5)}
\eea
Although all Young diagrams shown have a finite number of boxes, all row and column lengths would scale as $N$
as we take the large $N\to\infty$ limit.
In terms of these labels, the original loop operator is
\bea
O(\{ n_k\})=\sum_{T,(t,u)\alpha\beta}
\sqrt{f_T {\rm hooks}_t{\rm hooks}_u\over {\rm hooks}_T}\chi_{T,(t,u)\alpha\beta}(\sigma^{-1}_{\{n_k\}})
O_{T,(t,u)\beta\alpha}(Z,Y)
\eea
Our proposal for the operator localized in the LLM background $B$ is
\bea
O_B (\{ n_k\})=\sum_{T,(t,u)\alpha\beta}
\sqrt{f_{T_B} {\rm hooks}_{t_B}{\rm hooks}_u\over {\rm hooks}_{T_B}}\chi_{T_B,(t_B,u)\alpha\beta}(\sigma^{-1}_{\{n_k\}})
O_{T_B,(t_B,u)\beta\alpha}(Z,Y)
\eea
Some notation: $B\vdash n_B$, $T\vdash n_C+m$, $t\vdash n_C$ and $n=n_B+n_C$.
$n_B$ counts the number of $Z$ fields in the background; $n_C$ counts the number of $Z$ fields in the excitation.
Both sums above run over all Young diagrams $T\vdash m+n_C$, $t\vdash n_C$, $s\vdash m$ and the same multiplicity index.
We are interested in the limit in which $n_B\sim N^2$ and $n_C\sim\sqrt N$.
We want to compute the action of the dilatation operator on $O_B (\{ n_k\})$.
To accomplish this it is useful to simplify the above expression for $O_B (\{ n_k\})$.
The top most and left most box of $R$ is added to column $c$ and row $r$ of $B$.
In the above example $c=6$ and $r=8$; in the limit we consider both $r$ and $c$ are of order $N$.
Let $r_1$ denote the length of the first row of $B$ and let $c_1$ denote the length of the first column of $B$.
The first simplification we use comes from noticing that, at large $N$ we have
\bea
   {d_{T_B} n! m!\over d_{t_B} d_u (n+m)!}&=&{{\rm hooks}_{t_B}{\rm hooks}_{u}\over {\rm hooks}_{T_B}}\cr
     &=&\left({c_1-r\over c_1-r+c}{r_1-c\over r_1-c+r}\right)^m
      {{\rm hooks}_{t}{\rm hooks}_{u}\over {\rm hooks}_{T}}(1+O(N^{-1}))\cr
     &=&\kappa^m {{\rm hooks}_{t}{\rm hooks}_{u}\over {\rm hooks}_T}(1+O(N^{-1}))\label{hooksID}
\eea
The precise form of $\kappa$ does depend on the details of $B$.
However, this is the only source of $B$ dependence and the structure
\bea
   {d_{T_B} n! m!\over d_{t_B} d_u (n+m)!}=\kappa^m {{\rm hooks}_{t}{\rm hooks}_{u}\over {\rm hooks}_T}(1+O(N^{-1}))
\eea
holds for any $B$\footnote{Each corner in the Young diagram would be associated with a different value for $\kappa$.
For more than one excitation, at more than one corner, we'd have a product of terms, one for each corner.
The term for a given corner is the value of $\kappa$ for that corner raised to the power of the number of $Y$s 
appearing in the excitation at that corner.}.
The second simplification is in the value of the restricted character $\chi_{T_B,(t_B,u)\alpha\beta}(\sigma^{-1}_{\{n_k\}})$.
Here we will see that the fact that $\sigma^{-1}_{\{n_k\}}$ is an $n_C+m$ cycle plays a crucial role.
In terms of the original loop operator in the original AdS$_5\times$S$^5$ geometry, this is equivalent to the statement that
the loop operator is a single trace.
To compute the restricted character we will use a specific representation of the symmetric group called Young's orthogonal
representation\cite{hammer}.
The answer does not depend on the representation we use and we simply use this representation for convenience.
Young's orthogonal representation is defined by a rule which determines matrix elements of the matrices representing adjacent
permutations, that is, permutations of the form $(i,i+1)$. 
The adjacent permutations generate the complete group. 
The matrix elements for adjacent permutations are defined using the Young-Yamonouchi basis.
In this basis, each box in the Young diagram is assigned a unique number. 
In our conventions, if the boxes are removed removing box 1 first, box 2 second and so on, at each step one always obtains a
valid Young diagram. 
The collection of the labeled boxes is called a Young-Yamonouchi pattern, or more simply, just a pattern.
In addition to the integer assigned by the pattern, each box is associated with a second integer, called the content of the box.
The box in row $a$ and column $b$ has content $b-a$. 
The content of the box labeled $i$ in the pattern is denoted $c_i$.
If $(i,i+1)$ acts on a given state, it gives the same state back with coefficient ${1\over c_i-c_{i+1}}$ and it gives the state
labeled by the pattern with $i$ and $i+1$ swapped, with coefficient $\sqrt{1-{1\over (c_i-c_{i+1})^2}}$. 
Here is an example
\bea
\Gamma_{\tiny \yng(2,2,1)}\left( (12)\right)\ket{\young(53,42,1)} =
 -{1\over 2}\ket{\young(53,42,1)}
 +{\sqrt{3}\over 2}\ket{\young(53,41,2)}
\eea
Notice that if the difference in content between the two boxes scales as $N$ i.e. $c_i-c_{i+1}\sim N$, then at large $N$ the
permutation $(i,i+1)$ acting on a given pattern produces a linear combination of states, one with a pattern given by swapping 
boxes $i$  and $i+1$ in the original pattern (with coefficient $1+O(N^{-1})$) and one with the original pattern (with 
coefficient $O(N^{-1})$).
When we want to compute traces we are summing the diagonal elements of the representation of a permutation, so that we 
often need to keep the terms with the original pattern, which is suppressed at large $N$.
This observation was used to obtain the large $N$ limit of the one loop dilatation operator in \cite{Carlson:2011hy,Koch:2011hb}.
Using this representation, consider the computation of the restricted character 
$\chi_{T_B,(t_B,u)\alpha\beta}(\sigma^{-1}_{\{n_k\}})$, given by taking a (restricted) trace.
Recall that our goal is to write down the local versions of the restricted Schur polynomials given in (\ref{possSchurs}).
These are to be summed to produce a loop operator with three $Y$s and three $Z$s.
The positions of the first three boxes to be removed (these are associated with the $Y$s; they are the boxes that
are assembled to produce $s$) are fixed by the shapes of $R$ and $r$.
After removing these first three boxes, we need to label a further three boxes.
This follows because $\sigma^{-1}_{\{n_k\}}$ describes a loop with three $Y$s and three $Z$s and is thus a six-cycle.
All we need is the labels of the first six boxes
in the Young-Yamonouchi pattern if we are to evaluate the action of $\sigma^{-1}_{\{n_k\}}$ on the state.
We could remove them all from the vicinity of the first three boxes (see the diagram on the left below) or we could
include some more distant boxes (see the diagram on the right below for an example).
\bea
R_B={\tiny 
\young({\,}{\,}{\,}{\,}{\,}{\,}{\,}{\,}{\,}{\,}{\,}{\,}{\,}{\,},{\,}{\,}{\,}{\,}{\,}{\,}{\,}{\,}{\,}{\,}{\,}{\,}{\,}{\,},{\,}{\,}{\,}{\,}{\,}{\,}{\,}{\,}{\,}{\,}{\,}{\,}{\,}{\,},{\,}{\,}{\,}{\,}{\,}{\,}{\,}{\,}{\,}{\,}{\,}{\,}{\,}{\,},{\,}{\,}{\,}{\,}{\,}{\,}{\,}{\,}{\,}{\,}{\,}{\,}{\,}{\,},{\,}{\,}{\,}{\,}{\,}{\,}{\,}{\,}{\,}{\,}{\,}{\,}{\,}{\,},{\,}{\,}{\,}{\,}{\,}{\,}{\,}{\,}{\,}{\,}{\,}{\,}{\,}{\,},{\,}{\,}{\,}{\,}{\,}{*}{*}{2},{\,}{\,}{\,}{\,}{\,}{*}{1},{\,}{\,}{\,}{\,}{\,}{3},{\,}{\,}{\,}{\,}{\,},{\,}{\,}{\,}{\,}{\,})
}
\qquad\qquad\qquad
R_B={\tiny 
\young({\,}{\,}{\,}{\,}{\,}{\,}{\,}{\,}{\,}{\,}{\,}{\,}{\,}{\,},{\,}{\,}{\,}{\,}{\,}{\,}{\,}{\,}{\,}{\,}{\,}{\,}{\,}{\,},{\,}{\,}{\,}{\,}{\,}{\,}{\,}{\,}{\,}{\,}{\,}{\,}{\,}{\,},{\,}{\,}{\,}{\,}{\,}{\,}{\,}{\,}{\,}{\,}{\,}{\,}{\,}{\,},{\,}{\,}{\,}{\,}{\,}{\,}{\,}{\,}{\,}{\,}{\,}{\,}{\,}{\,},{\,}{\,}{\,}{\,}{\,}{\,}{\,}{\,}{\,}{\,}{\,}{\,}{\,}{\,},{\,}{\,}{\,}{\,}{\,}{\,}{\,}{\,}{\,}{\,}{\,}{\,}{\,}{*},{\,}{\,}{\,}{\,}{\,}{\,}{\,}{2},{\,}{\,}{\,}{\,}{\,}{\,}{1},{\,}{\,}{\,}{\,}{\,}{3},{\,}{\,}{\,}{\,}{\,},{\,}{\,}{\,}{*}{*})
}
\eea
The important observation is that only states of the form given in the diagram on the left above contribute.
Concretely, after acting with $\sigma^{-1}_{\{n_k\}}$ on the states of the form given in the diagram on the right above, we
always find (at the leading order in large $N$) that some of the first $m$ labels (given in our example by 1,2,3) are transported
into the distant boxes.
This follows from the fact that (i) the difference in content between the distant boxes and the first $m$ boxes is order $N$ and
hence (ii) in Young's orthogonal representation, transpositions between the first $m$ and the remaining boxes always transport
local boxes to the location of the distant boxes.
These states will not contribute to the trace because the overlap of the state with the original box locations and the state
with some local boxes swapped with distant boxes vanishes.
Thus, the trace only receives contributions from states of the form given in the diagram on the left above.
The contribution from the states of the form given in the diagram on the right above are of order $N^{-1}$.
In general, the $m+n_C$ cycle ``ties'' the $n_C+m$ labeled boxes together.
This forces the $n_C$ $Z$ boxes (which might have appeared in any distant corner of $R_B$) to sit adjacent to the $Y$ boxes.
At this stage it is useful to introduce a bijection between states in $R$ and subspaces in $R_B$ as follows
\bea
   {\tiny 
\young({\,}{\,}{\,}{\,}{\,}{\,}{\,}{\,}{\,}{\,}{\,}{\,}{\,}{\,},{\,}{\,}{\,}{\,}{\,}{\,}{\,}{\,}{\,}{\,}{\,}{\,}{\,}{\,},{\,}{\,}{\,}{\,}{\,}{\,}{\,}{\,}{\,}{\,}{\,}{\,}{\,}{\,},{\,}{\,}{\,}{\,}{\,}{\,}{\,}{\,}{\,}{\,}{\,}{\,}{\,}{\,},{\,}{\,}{\,}{\,}{\,}{\,}{\,}{\,}{\,}{\,}{\,}{\,}{\,}{\,},{\,}{\,}{\,}{\,}{\,}{\,}{\,}{\,}{\,}{\,}{\,}{\,}{\,}{\,},{\,}{\,}{\,}{\,}{\,}{\,}{\,}{\,}{\,}{\,}{\,}{\,}{\,}{\,},{\,}{\,}{\,}{\,}{\,}{6}{4}{2},{\,}{\,}{\,}{\,}{\,}{5}{1},{\,}{\,}{\,}{\,}{\,}{3},{\,}{\,}{\,}{\,}{\,},{\,}{\,}{\,}{\,}{\,})
}\qquad \leftrightarrow\qquad{\tiny \young(642,51,3)}\label{newstates}
\eea
Each of these subspaces is an irrep of $S_{n_B}$ equivalent to the irrep labeled by $B$.
This map is equivariant\footnote{This follows from the fact that the action of a group element depends only on the differences of the
content in the labeled boxes, and these differences are equal for the two states appearing in (\ref{newstates}).} with respect to 
the action of the cycle $\sigma^{-1}_{\{n_k\}}$ so that we now find
\bea
   \chi_{T_B,(t_B,u)\alpha\beta}(\sigma^{-1}_{\{n_k\}})=d_B\chi_{T,(t,u)\alpha\beta}(\sigma^{-1}_{\{n_k\}})
\eea
where $d_B$ is the dimension of symmetric group representation $B$.
This result holds only in the large $N$ limit.

It is useful to introduce notation with which to describe the different vector spaces that enter into our analysis. 
The states labeled by patterns filling $R_B$ are denoted $|R_B,a\rangle$ with $a=1,...,d_{R_B}$.
The states of the type shown in (\ref{newstates}) are written as $|\hat{R}_B,a\rangle$ with $a=1,...,d_Rd_B$
and $|R,a\rangle$ with $a=1,...,d_{R}$ respectively.
The result we have found above is written as
\bea
   \chi_{T_B,(t_B,u)\alpha\beta}(\sigma^{-1}_{\{n_k\}})=\chi_{\hat{T}_B,(\hat{t}_B,u)\alpha\beta}(\sigma^{-1}_{\{n_k\}})
   =d_B\chi_{T,(t,u)\alpha\beta}(\sigma^{-1}_{\{n_k\}})
\eea
with the new notation.
The intermediate step makes it explicit that only certain states contribute to the restricted trace.
The first equality above is only true at large $N$; the second is exact.

The above discussion has shown that we can restrict the states participating in the trace; from now on we will
restrict all traces in this way - it simplifies our discussion of the action of the dilatation operator dramatically.
Using these simplifications, we can write our proposal for the operator localized in the LLM background $B$ as
\bea
O_B (\{ n_k\})
&=&\sum_{T,(t,u)\alpha\beta}
\sqrt{f_{T_B} {\rm hooks}_{t_B}{\rm hooks}_u\over {\rm hooks}_{T_B}}
\chi_{\hat{T}_B,(\hat{t}_B,u)\alpha\beta}(\sigma^{-1}_{\{n_k\}})
O_{\hat{T}_B,(\hat{t}_B,u)\beta\alpha}(Z,Y)\cr
&&\!\!\!\!\!\!\!\!\!\!\!\!\!\!\!\!\!\!\!\!
=\kappa^{m\over 2} d_B \sum_{T,(t,u)\alpha\beta}
\sqrt{f_{T_B} {\rm hooks}_{t}{\rm hooks}_u\over {\rm hooks}_{T}}
\chi_{T,(t,u)\alpha\beta}(\sigma^{-1}_{\{n_k\}})
O_{\hat{T}_B,(\hat{t}_B,u)\beta\alpha}(Z,Y)
\eea
This looks remarkably similar to the original local loop operator
\bea
O(\{ n_k\})=\sum_{T,(t,u)\alpha\beta}
\sqrt{f_{T} {\rm hooks}_{t}{\rm hooks}_u\over {\rm hooks}_{T}}
\chi_{T,(t,u)\alpha\beta}(\sigma^{-1}_{\{n_k\}})
O_{T,(t,u)\beta\alpha}(Z,Y)
\eea

Naively one may have expected to treat all the $Z$ fields in our operator, on the same footing. 
The arguments of this section motivate the fact that for the operators we propose, this is not the case.
The reason why some of the $Z$s are treated differently is quite transparent in our analysis: the 
permutation $\sigma^{-1}_{\{n_k\}}$ has tied some of the $Z$ fields with the $Y$ fields. 
The $Y$s are localized on the Young diagram, so that the $Z$s tied to the $Y$s will be localized to this region too.
This is rather natural as these fields are supposed to constitute a single object: the closed string.
Recall that the number of $Z$s in the closed string is $n_C$ and the number of $Z$s in the background is $n_B$.
Thanks to the permutation $\sigma^{-1}_{\{n_k\}}$, the first $m+n_C$ labeled boxes are localized on $R_B$.
$r_B$ is filled with patterns that label states in an irreducible representation of  $S_{n_C+n_B}=S_n$ while $\hat{r}_B$ is filled
with patterns that label states in an irreducible representation of $S_{n_C}\times S_{n_B}$.
These are two very different things.
For further discussion of these localized restricted Schur polynomials, see Appendix \ref{localrschur}.

\section{LLM Magnons}\label{LLMmag}

In this section we would like to evaluate the action of the one loop dilatation operator on our proposed localized loops 
in the LLM background $B$.
Our goal is to compute the one loop anomalous dimension of the localized loop, a quantity that can be
compared to energies in the dual string theory, to either support or rule out our proposal.
To begin we will compute $DO_{\hat{R}_B,(\hat{r}_B,s)\mu_1\mu_2}(Z,Y)$ which is needed for the evaluation
of $DO_B(\{n_k\})$ - the object of interest to us.
Using the exact one loop result, we find\footnote{The same result would be obtained by acting on 
$O_{R_B,(r_B,s)\mu_1\mu_2}(Z,Y)$ and using the simplifications of large $N$.}
\bea
  DO_{\hat{R}_B,(\hat{r}_B,s)\mu_1\mu_2}(Z,Y)
=\sum_{T,(t,u)\nu_1\nu_2}N_{\hat{R}_B,(\hat{r}_B,s)\mu_1\mu_2;T,(t,u)\nu_1\nu_2}O_{T,(t,u)\nu_1\nu_2}(Z,Y)
\label{dilact}
\eea
where
\bea
N_{\hat{R}_B,(\hat{r}_B,s)\mu_1\mu_2;T,(t,u)\nu_1\nu_2}
=-{g_{YM}^2\over 8\pi^2}\sum_{\hat{R}_B'}
{c_{R_B,R_B'} d_T n m \over d_{\hat{R}_B'}d_t d_u (n+m)}
\sqrt{f_T {\rm hooks}_T{\rm hooks}_{r_B}{\rm hooks}_s\over f_{R_B}{\rm hooks}_{R_B}
{\rm hooks}_t{\rm hooks}_u}\cr
{\rm Tr}\left(\big[ (1,m+1),P_{\hat{R}_B,(\hat{r}_B,s)\mu_1\mu_2}\big]I_{R_B'T'}
\big[(1,m+1),P_{T,(t,u)\nu_2\nu_1}\big]I_{T'R_B'}\right)\label{matelem}
\eea
A few comments are in order. 
The one loop dilatation operator action was derived when acting on a restricted Schur polynomial,
$O_{R_B,(r_B,s)\mu_1,\mu_2}(Z,Y)$.
Above we are acting on $O_{\hat{R}_B,(\hat{r}_B,s)\mu_1\mu_2}(Z,Y)$. 
By thinking of $O_{\hat{R}_B,(\hat{r}_B,s)\mu_1\mu_2}(Z,Y)$ as a linear combination of restricted Schur polynomials,
it is straightforward to repeat the derivation given in \cite{DeComarmond:2010ie}.
There are two possible swaps that contribute.
The first swap is of the form $(1,m+1)$ which acts on a $Y$ slot and a $Z$ slot which belongs to the localized loop; this is the
permutation that appears in (\ref{dilact}). 
The second swap is of the form $(1,m+n_C+1)$ and it acts on a $Y$ slot and a $Z$ slot which belongs to the background; it
gives a contribution that is suppressed in the large $N$ limit and so we drop it.
This follows from the fact that $m+n_C+1$ must appear in a corner, and after stripping off the first $m+n_C$ boxes, the
only corners remaining are distant.
In this way we learn that (\ref{dilact}) is not exact: use of the large $N$ limit has been made to discard a specific interaction 
between the background and the loop.

To make further progress, we need to study the matrix elements $N_{\hat{R}_B,(\hat{r}_B,s)\mu_1\mu_2;T,(t,u)\nu_1\nu_2}$.
Our first task is to characterize the labels $T,(t,u)\nu_2\nu_1$ of operators that contribute in (\ref{dilact}).
The intertwiner $I_{T'R_B'}$ is only non-zero when $R_B$ and $T$ differ by the placement of at 
most one box - it is only in this case that a non-zero map can be defined.
To proceed further, we need a little more notation. 
Specifically, we need to spell out which slots in the restricted Schur polynomial are associated to which representations. 
Writing the restricted Schur polynomial as
\bea
O_{\hat{R}_B,(\hat{r}_B,s)\mu_1\mu_2}(Z,Y)
={1\over (n+m)!}\sum_{\sigma\in S_{n+m}}
\chi_{\hat{R}_B,(\hat{r}_B,s)\mu_1\mu_2}(\sigma)
Y^{i_1}_{i_{\sigma(1)}}\cdots Y^{i_m}_{i_{\sigma(m)}}\cr
Z^{i_{m+1}}_{i_{\sigma(m+1)}}\cdots
Z^{i_{m+n_C}}_{i_{\sigma(m+n_C)}}Z^{i_{m+n_C+1}}_{i_{\sigma(m+n_C+1)}}\cdots
Z^{i_{m+n}}_{i_{\sigma(m+n)}}
\eea
associates the first $m+n_C$ slots to the closed string excitation and the last $n_B$ slots to the background.
We will now make use of the Casimirs of the symmetric group, given by summing all the elements in a given conjugacy class.
The symmetric groups and Casimirs which play a role are
\begin{itemize} 
\item[1.] $S_{n_B+n_C-1}$ which permutes the indices $m+2,m+3,\cdots,m+n$ has Casimirs ${\cal C}_i^{n_B+n_C-1}$.
               Recall that $n=n_B+n_C$. 
\item[2.] $S_{n_B}$ which permutes the indices $m+n_C+1,m+n_C+2,\cdots,m+n$  has Casimirs ${\cal C}_i^{n_B}$.
\item[3.] $S_{n_C-1}$ which permutes the indices $m+2,m+3,\cdots,m+n_C$  has Casimirs ${\cal C}_i^{n_C-1}$.
\item[4.] $S_{m-1}$ which permutes the indices $2,3,\cdots,m$  has Casimirs ${\cal C}_i^{m-1}$.
\end{itemize}
These Casimirs distinguish a representation.
Knowing the value of the Casimir associated to a given conjugacy class when acting on any state belonging to a particular 
representation is equivalent to knowing the value of the character in the representation for the relevant conjugacy class.
Knowing the value of all the Casimirs is equivalent to knowing the complete set of characters, which specifies the representation
completely.
Denoting the state labeled by $a$ in some representation $R$ (not necessarily an irrep) we have
\bea
   {\cal C}_i |R,a\rangle =\lambda^R_i |R,a\rangle
\eea
We get the same value of $\lambda^R_i$ no matter what state (i.e. what value of $a$) we act on and knowing the
complete set of $\lambda^R_i$s allows us to determine $R$ completely, up to equivalence. 
Consider the trace we wish to compute
\bea
{\rm Tr}\left(\big[ (1,m+1),P_{\hat{R}_B,(\hat{r}_B,s)\mu_1\mu_2}\big]I_{R_B'T'}
\big[(1,m+1),P_{T,(t,u)\nu_2\nu_1}\big]I_{T'R_B'}\right)\cr
={\rm Tr}\left((1,m+1)P_{\hat{R}_B,(\hat{r}_B,s)\mu_1\mu_2}I_{R_B'T'}
(1,m+1),P_{T,(t,u)\nu_2\nu_1}I_{T'R_B'}\right)+...
\label{tracetodo}
\eea
There are three more terms in the dots above.
The projection operator $P_{\hat{R}_B\to (\hat{r}_B,s)\mu_1\mu_2}$ has the following form
\bea
P_{\hat{R}_B,(\hat{r}_B,s)\mu_1\mu_2}=\sum_i |i\rangle\langle i|
\eea
where the states $|i\rangle$ are linear combinations of states labeled by patterns with
all states having exactly the same boxes filled with all integers from $m+1$ to $m+n$.
It is thus possible to decompose this projector into a sum of terms (denoted $P^{a}_{\hat{R}_B,(\hat{r}_B,s)\mu_1\mu_2}$)
which each have their $m+1$ th box in a specific corner (with label $a$) of $\hat{r}_B$.
Denote the representation obtained by removing this corner box from $r_B$ by $r_B'(a)$.
In this case we can write
\bea
P_{\hat{R}_B,(\hat{r}_B,s)\mu_1\mu_2}=\sum_a P^{a}_{\hat{R}_B,(\hat{r}_B,s)\mu_1\mu_2}\cr
{\cal C}_i^{n_B+n_C-1}P^{a}_{\hat{R}_B,(\hat{r}_B,s)\mu_1\mu_2} =
\lambda_i^{r_B(a)'} P^{a}_{\hat{R}_B,(\hat{r}_B,s)\mu_1\mu_2}\label{forr}
\eea
A similar decomposition gives
\bea
P_{T,(t,u)\nu_2\nu_1}=\sum_a P_{T,(t,u)\nu_2\nu_1}^a\cr
{\cal C}_i^{n_B+n_C-1}P_{T,(t,u)\nu_2\nu_1}^a =
\lambda_i^{t(a)'}P_{T,(t,u)\nu_2\nu_1}^a\label{fort}
\eea

Let us now show a concrete example of how these Casimirs can be used to derive restrictions on the labels
$T,(t,u)$ given the labels $\hat{R}_B,(\hat{r}_B,s)$.
For each of the four terms in (\ref{tracetodo}) we can argue as follows
\bea
\lambda_i^{r_B(a)'}{\rm Tr}\left((1,m+1)P^{a}_{\hat{R}_B,(\hat{r}_B,s)\mu_1\mu_2}I_{R_B'T'}
(1,m+1)P^{b}_{T,(t,u)\nu_2\nu_1}I_{T'R_B'}\right)\cr
={\rm Tr}\left((1,m+1)P^{a}_{\hat{R}_B,(\hat{r}_B,s)\mu_1\mu_2}{\cal C}_i^{n_B+n_C-1} I_{R_B'T'}
(1,m+1)P^{b}_{T,(t,u)\nu_2\nu_1}I_{T'R_B'}\right)\cr
={\rm Tr}\left((1,m+1)P^{a}_{\hat{R}_B,(\hat{r}_B,s)\mu_1\mu_2} I_{R_B'T'}(1,m+1)
{\cal C}_i^{n_B+n_C-1} P^{b}_{T,(t,u)\nu_2\nu_1}I_{T'R_B'}\right)\cr
=\lambda_i^{t(b)'}{\rm Tr}\left((1,m+1)P^{a}_{\hat{R}_B,(\hat{r}_B,s)\mu_1\mu_2} I_{R_B'T'}
 (1,m+1)P^{b}_{T,(t,u)\nu_2\nu_1}I_{T'R_B'}\right)
\eea
The first equality uses (\ref{forr}).
The second equality is a consequence of the fact that
\bea
\big[ {\cal C}_i^{n_B+n_C-1} ,I_{\hat{R}_B'T'} (1,m+1)\big]=0
\eea
which follows because $(1,m+1)$ commutes with all elements of $S_{n_B+n_C-1}$ and the intertwiner maps between
equivalent representations of  $S_{n_B+n_C-1}$.
The final equality uses (\ref{fort}).
Thus, we have learned that if $\lambda_i^{r_B(a)'}\ne \lambda_i^{t(b)'}$, or equivalently if $r_B(a)'\ne t(b)'$ we have
\bea
{\rm Tr}\left((1,m+1)P^{a}_{\hat{R}_B,(\hat{r}_B,s)\mu_1\mu_2} I_{R_B'T'}(1,m+1)
 P^{b}_{T,(t,u)\nu_2\nu_1}I_{T'R_B'}\right)=0
\eea
Now, writing
\bea
{\rm Tr}\left((1,m+1)P_{\hat{R}_B,(\hat{r}_B,s)\mu_1\mu_2} I_{R_B'T'}(1,m+1)
 P_{T,(t,u)\nu_2\nu_1}I_{T'R_B'}\right)\cr
=\sum_a\sum_b
{\rm Tr}\left((1,m+1)P^{a}_{\hat{R}_B,(\hat{r}_B,s)\mu_1\mu_2} I_{R_B'T'}(1,m+1)
 P^{b}_{T,(t,u)\nu_2\nu_1}I_{T'R_B'}\right)
\eea
we learn that the trace vanishes unless $r_B$ and $t$ differ by at most the placement of a single box.
We can refine this analysis further.
In constructing the operator $O_{\hat{R}_B,(\hat{r}_B,s)\mu_1\mu_2}(Z,Y)$, we know that not all states in $r_B$
participate: only the states in the irrep $(r,B)$ of $S_{n_C}\times S_{n_B}$ participate.
Our shorthand for ``states in the irrep $(r,B)$ of $S_{n_C}\times S_{n_B}$'' is $\hat{r}_B$.
We will demonstrate that $t$ also belongs to an $S_{n_C}\times S_{n_B}$ representation.
We know that
\bea
{\cal C}_i^{n_B}P_{\hat{R}_B,(\hat{r}_B,s)\mu_1\mu_2} =
\lambda_i^{B} P_{\hat{R}_B,(\hat{r}_B,s)\mu_1\mu_2}\label{forR}
\eea
Now, decompose irreducible representation $t$ in $P_{T,(t,u)\nu_2\nu_1}$ into a direct sum of irreducible
representations of the $S_{n_B}$ subgroup as follows
\bea
 P_{T,(t^{b},u)\nu_2\nu_1}=\sum_b  P^{b}_{T,(t^{b},u)\nu_2\nu_1}
\eea
Arguing exactly as we did above we learn that the trace is only non-zero if $t^b=B$. This implies that
we can replace $T$ and $t$ by $\hat T_B$ and $\hat t_b$ respectively.
Finally, using the Casimirs ${\cal C}_i^{n_C-1}$ and ${\cal C}_i^{m-1}$ we can argue that $r$ and $t$ as well $s$ and $u$
differ by at most the placement of a single box.
So, the mixing problem is tightly constrained: operators $O_{\hat{R}_B,(\hat{r}_B,s)\mu_1\mu_2}(Z,Y)$
(using irrep $(B,r,s)$ of the subgroup $S_{n_B}\times S_{n_C}\times S_m$) only mix with operators
$O_{\hat{T}_B,(\hat{t}_B,u)\nu_1\nu_2}(Z,Y)$ (using irrep $(B,t,u)$ of the subgroup) if the Young diagrams in the pairs
$s,u$, as well as $r,t$ and $R,T$ differ by at most the placement of a single box. 

A comment is in order.
In the previous section we have proposed operators dual to local excitations on some background geometry.
Our proposal puts the fields belonging to the background into one representation and the fields belonging to the
excitation into a second representation.
The result we have obtained above shows that the dilatation operator, which is to be identified with the Hamiltonian
of the excitation, respects this structure: only operators that have this same make-up, with the same representation
for the background, mix.
This is support in favor of our construction.

Since $T_B$ and $R_B$ can differ by at most one block we need to consider the following possibility
\bea
R_B={\tiny \yng(14,14,14,14,14,14,14,8,7,6,5,5)}
\qquad
T_B={\tiny \yng(15,14,14,14,14,14,14,8,7,5,5,5)}
\eea
i.e. $T_B$ has a distant box added. If this is a $Y$ box, we will have
\bea
t_B={\tiny \yng(14,14,14,14,14,14,14,7,6,5,5,5)}
\eea
i.e. $t_B$ has no distant boxes. 
Since the box we must drop is a $Y$ box, the swap $(1,m+1)$ appearing in (\ref{tracetodo}) must leave the boxes inert, 
and this immediately implies that this dilatation operator matrix element is of order $N^{-1}$.
This proves that under the action of the dilatation operator, none of the $Y$ boxes are transported to distant locations
on the Young diagram.
Since the distant box in $T_B$ is a $Z$ box, we will have
\bea
t_B={\tiny \yng(15,14,14,14,14,14,14,6,6,5,5,5)}
\eea
We have now learned enough about the matrix elements 
$N_{\hat{R}_B,(\hat{r}_B,s)\mu_1\mu_2;\hat{T}_B,(\hat{t}_B,u)\nu_1\nu_2}$
that we can consider the action of the dilatation operator on a localized loop
\bea
D O_B(\{ n_k\})
&=&\sum_{R,(r,s)\mu_1\mu_2}
\sqrt{f_{R_B} {\rm hooks}_{r_B}{\rm hooks}_u\over {\rm hooks}_{R_B}}
\chi_{\hat{R}_B,(\hat{r}_B,s)\mu_2\mu_1}(\sigma^{-1}_{\{n_k\}})
D O_{\hat{R}_B,(\hat{r}_B,s)\mu_1\mu_2}(Z,Y)\cr
&=&\sum_{R,(r,s)\mu_1\mu_2}\sum_{T,(t,u)\nu_1\nu_2}
\sqrt{f_{R_B} {\rm hooks}_{r_B}{\rm hooks}_u\over {\rm hooks}_{R_B}}
\chi_{\hat{R}_B,(\hat{r}_B,s)\mu_2\mu_1}(\sigma^{-1}_{\{n_k\}})\cr
&&\times N_{\hat{R}_B,(\hat{r}_B,s)\mu_1\mu_2;\hat{T}_B,(\hat{t}_B,u)\nu_1\nu_2}
O_{\hat{T}_B,(\hat{t}_B,u)\nu_1\nu_2}(Z,Y)\cr
&=&-{g_{YM}^2\over 8\pi^2}\sum_{p=1}^m\sum_{q=m+1}^{m+n}
\sum_{T^+,t,u,\nu_1\nu_2}\sum_T c_{T^+ T}{{\rm hooks}_T\over{\rm hooks}_{T^+}}
\sqrt{f_{T_B} {\rm hooks}_{t_B}{\rm hooks}_u\over {\rm hooks}_{T_B}}\cr
&&\qquad\times\chi_{T^+,(t,u,{\tiny\yng(1)})\nu_1\nu_2}(\psi^{-1})
O_{\hat{T}_B,(\hat{t}_B,u)\nu_2\nu_1}(Z,Y)\label{finres}
\eea
where to get the last line above we have made use of the identity proved in Appendix \ref{identities}.
Focus on the character $\chi_{T^+,(t,u,{\tiny\yng(1)})\nu_1\nu_2}(\psi^{-1})$.
There is an extra box in $T^+$ that must be dropped to obtain $T$.
The permutation $\psi$ depends on $p$ and $q$.
If the permutation $\psi^{-1}$ consists of cycles that mix the indices of the local fields with the possible
distant $Z$ box, then we know by the arguments of section \ref{LocalizedString} that these contributions 
can be dropped at large $N$.
In Appendix \ref{identities} we argue that $\psi^{-1}$ has two cycles, one of length $k$ and one of length $n_C+m+1-k$. 
The distant box will not be suppressed as long as it appears in a cycle that does not tie it to any local boxes.
There are only two such terms.
At large $N$, there are $mn_C=O(N)$ terms appearing on the right hand side of (\ref{finres}), so that this is a subleading
contribution.
This proves that at large $N$, any terms with a distant $Z$ box can be neglected.
Our argument has demonstrated that only local loop operators, as we have defined them, contribute.
This is enough to prove that
\bea
{\rm Tr}\left(\big[ (1,m+1),P_{\hat{R}_B,(\hat{r}_B,s)\mu_1\mu_2}\big]I_{R_B'T_B'}
\big[(1,m+1),P_{\hat T_B,(\hat t_B ,u)\nu_2\nu_1}\big]I_{T_B'R_B'}\right)\cr
=d_B
{\rm Tr}\left(\big[ (1,m+1),P_{R,(r,s)\mu_1\mu_2}\big]I_{R'T'}
\big[(1,m+1),P_{T,(t,u)\nu_2\nu_1}\big]I_{T'R'}\right)\label{frstid}
\eea
To prove the equality note that the map $I_{T_B' R_B'}$ has a trivial action: it simply maps between two subspaces.
It is the permutation $(1,m+1)$ that has a nontrivial action.
However, as we have discussed, there is a map between $\hat{R}_B$ and $R$ that is equivariant with respect to the action
of $(1,m+1)$.
The only difference between the two sides of (\ref{frstid}) is that the LHS has a contribution from each possible pattern
of the background Young diagram $B$.
This is the origin of $d_B$ on the right hand side.
Apart from the above trace, the matrix elements $N_{\hat{R}_B,(\hat{r}_B,s)\mu_1\mu_2;\hat T_B,(\hat t_B,u)\nu_1\nu_2}$
involve a few other factors.
Looking back at (\ref{matelem}), we will need the values of
\bea
  {d_{\hat{T}_B}\over d_{\hat{R}_B'}d_{\hat{t}_B} d_u}=   {d_T\over d_{R'}d_{t} d_u d_B}
\eea
\bea
   {n_C\over n_C+m}=1+O\left({m\over n_C}\right)={n\over n+m}
\eea
\bea
{{\rm hooks}_{T_B}{\rm hooks}_{r_B}{\rm hooks}_s\over {\rm hooks}_{R_B}
{\rm hooks}_{t_B}{\rm hooks}_u}=
\kappa^m {{\rm hooks}_{r}{\rm hooks}_s\over {\rm hooks}_R}
\kappa^{-m}{{\rm hooks}_T\over {\rm hooks}_t{\rm hooks}_u}=
{{\rm hooks}_{T}{\rm hooks}_{r}{\rm hooks}_s\over {\rm hooks}_{R}
{\rm hooks}_{t}{\rm hooks}_u}\cr
\eea
In the first formula above we used the fact that the first box dropped from $d_{\hat{R}_B'}$ is a $Y$ box, i.e.
it does not belong to $B$.
In the second equation above we used the fact that we have a dilute magnon gas (i.e. $m\ll n_C$).
In the last identity above we have used (\ref{hooksID}).
Finally, the only factors which don't cancel in the ratio ${f_{T_B}\over f_{R_B}}$ are the factors of boxes which
are not common between $T_B$ and $R_B$.
Denote the factor of the box labeled $n_C+m$ in the Young-Yamonouchi pattern for the states in $\hat R_B$ by $N_{\rm eff}$.
The values of ${f_{T_B}\over f_{R_B}}$ and $c_{R_B R_B'}$ are given by replacing $N$ in
${f_{T}\over f_{R}}$ and $c_{R R'}$ by $N_{\rm eff}$. 
This now proves that
\bea
N_{\hat{R}_B,(\hat{r}_B,s)\mu_1\mu_2;\hat T_B,(\hat t_B,u)\nu_1\nu_2}
=N_{R,(r,s)\mu_1\mu_2; T,(t_,u)\nu_1\nu_2}\Big|_{N\to N_{\rm eff}}
\eea
In the end, this is a remarkably simple result: the matrix elements of the dilatation operator acting on localized loops in the
LLM geometry are given by replacing $N\to N_{\rm eff}$ in the matrix elements of the dilatation operator in the trivial background!  
This provides a generalization of the ${1\over 2}$-BPS result proved in \cite{deMelloKoch:2009jc}.
Performing this replacement we now find that

\bea
 &&  DO_B(\{ n_k\})\cr
&=&{g_{YM}^2 N_{\rm eff}\over 8\pi^2}
\left(2O_B(\{n_k\})-O_B(\{n_1+1,n_2-1,\cdots,n_m\})-O_B(\{n_1-1,n_2+1,\cdots,n_m\})
\right)\cr
&+&{g_{YM}^2 N_{\rm eff}\over 8\pi^2}
\left(2O_B(\{n_k\})-O_B(\{n_1,n_2+1,n_3-1,\cdots,n_m\})-O_B(\{n_1,n_2-1,n_3+1,\cdots,n_m\})
\right)\cr
&&\quad \vdots
\cr
&+&{g_{YM}^2 N_{\rm eff}\over 8\pi^2}
\left(2O_B(\{n_k\})-O_B(\{n_1-1,n_2,\cdots,n_m+1\})-O_B(\{n_1+1,n_2,\cdots,n_m-1\})
\right)\cr
&&
\eea
We can diagonalize this action exactly as we do it in the planar limit: by Fourier transforming to momentum space.
For example, a state with two magnons, one of momentum $p$ and one of momentum $-p$ is given by
\bea
   O_{B}(p,-p)=\sum_{n_1=0}^{n_C}O_B(\{n_1,n_C-n_1\})e^{in_1 p}
\eea
\bea
   DO_{B}(p,-p)=(E(p)+E(-p))O_{B}(p,-p)\cr\cr
 E={g_{YM}^2 N\over 8\pi^2}{N_{\rm eff}\over N}\left(2\sin {p\over 2}\right)^2
   \equiv {N_{\rm eff}\over N} g^2 \left(2\sin {p\over 2}\right)^2\label{gtres}
\eea
This diagonalization and its perfect agreement with a string worldsheet description has been discussed many times in
the literature (see \cite{Constable:2002hw} for example).

\begin{figure}
\begin{center}
\includegraphics[height=7cm,width=9cm]{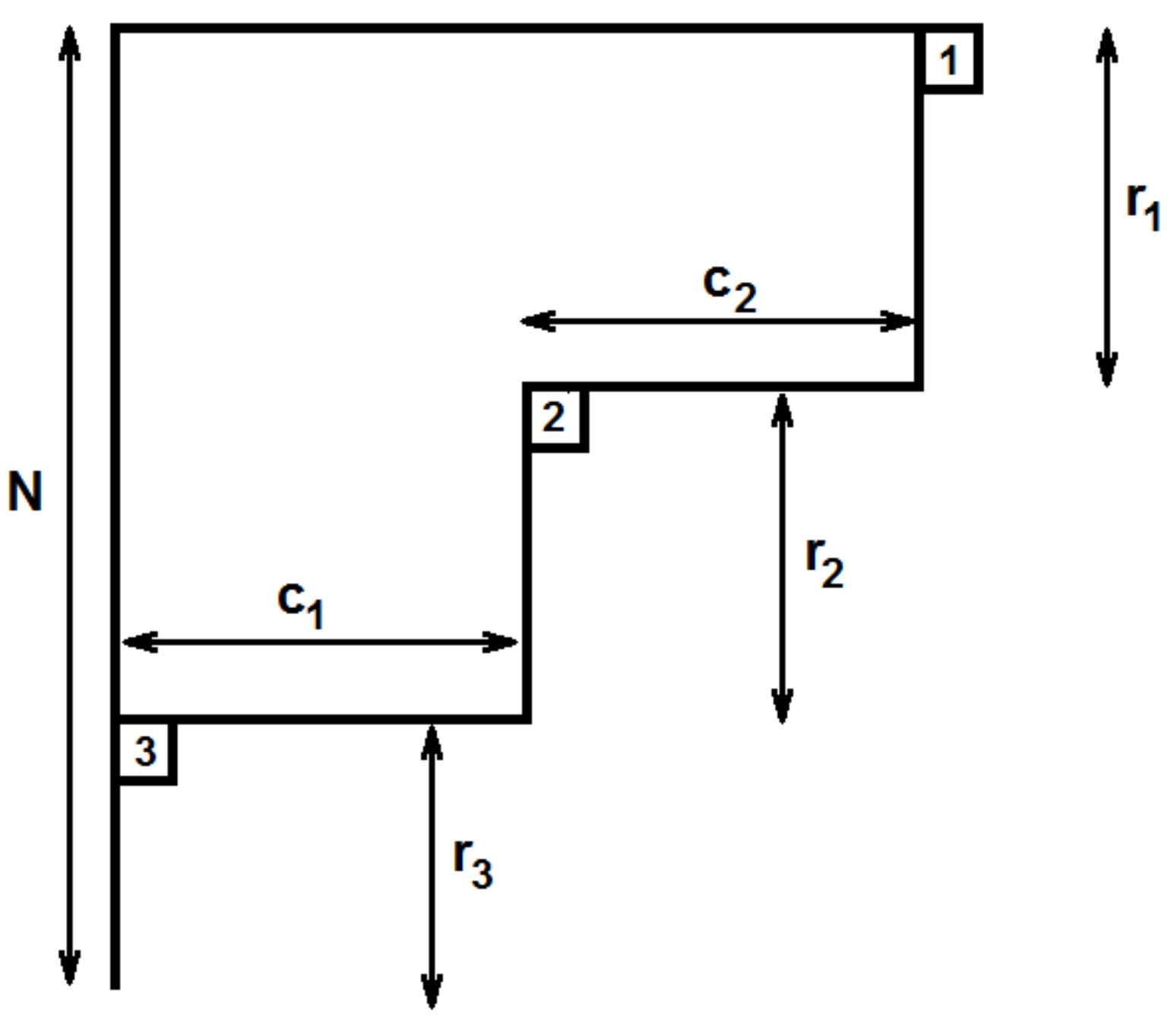}
\caption{A Young diagram $B$ which corresponds to the LLM background with boundary condition shown in Fig \ref{fig:LLMrings}.
The number of rows and columns defining $B$ as well as $N$ are shown.} 
\label{fig:labledYD} 
\end{center}
\end{figure}

Now that we have the gauge theory anomalous dimension associated to a magnon, we would like to compare it to the prediction 
from the string theory analysis.
We will compare the gauge theory and the string theory for localized loops located in the three regions possible.
For the gauge theory prediction we need the value of $N_{\rm eff}$ which is given by the factors of the boxes labeled in
Fig \ref{fig:labledYD}.
The factor of the box labeled $i$ is denoted $N_{\rm eff}^{(i)}$ with
\bea
   N_{\rm eff}^{(1)}=N+c_1+c_2,\qquad N_{\rm eff}^{(2)}=N+c_1-r_1,\qquad N_{\rm eff}^{(3)}=N-r_1-r_2
\eea
Noting that $N=r_1+r_2+r_3$ these can also be written as
\bea
   N_{\rm eff}^{(1)}=r_1+r_2+r_3+c_1+c_2,\qquad N_{\rm eff}^{(2)}=r_2+r_3+c_1,\qquad 
N_{\rm eff}^{(3)}=r_3
\eea
These factors have a very natural geometrical interpretation in the dual gravity.
The simplest way to map between the LLM boundary condition and the Young diagram describing the background is through
the use of the free fermion language\cite{Lin:2004nb}.
The LLM boundary condition is a picture of the phase space of $N$ non-interacting fermions in an external harmonic oscillator
potential. 
The central disk is a number of fermions (set by the area of the central disk divided by $\pi/N$) that have not been excited.
The inner black annulus is some number of fermions (set by the area of this annulus divided by $\pi/N$) each excited by the same 
amount (set by the area of the inner white annulus divided by $\pi/N$).
Finally, the outer black annulus is some number of fermions (set by the area of this annulus divided by $\pi/N$), with each again
excited by the same amount (set by the area of the outer white annulus divided by $\pi/N$).
The Young diagram is a picture of the same thing. 
Each row corresponds to a fermion.
The number of boxes in any given row is equal to the amount by which this fermion is excited.
Using this dictionary, we see that the central disk has an area of $\pi r_3/N$ and hence a radius squared of $r_3/N$.
Similarly, the inner black annulus has an outer radius squared of $(r_3+c_1+r_2)/N$ while the outer black annulus has an 
outer radius squared of $(r_1+r_2+r_3+c_1+c_2)/N$.
Thus, the radius (squared) on the LLM plane at which the loop is localized is set by the factor of the box divided by $N$.

\begin{figure}
\begin{center}
\includegraphics[height=5cm,width=5cm]{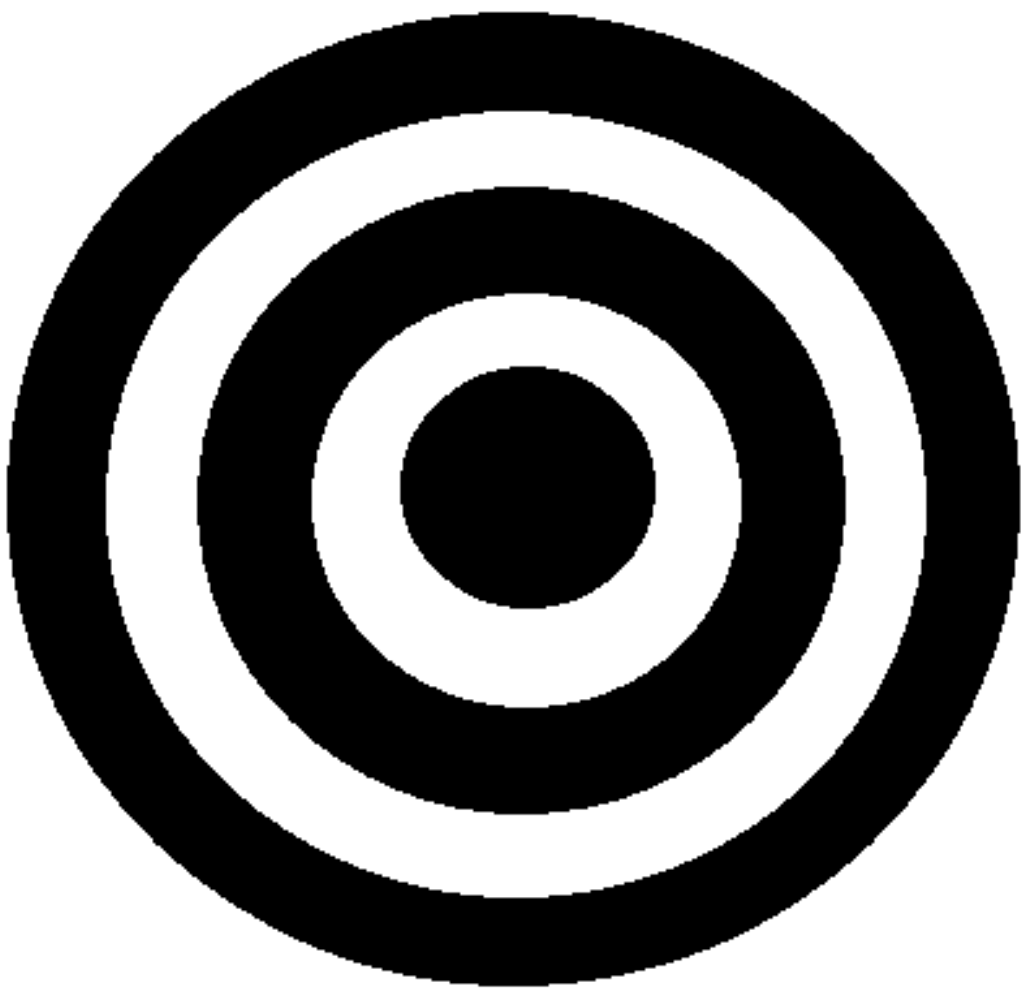}
\caption{This is the LLM boundary condition corresponding to the Young diagram given in Fig \ref{fig:labledYD}. The central black
disk has area equal to $\pi r_3$ and hence a radius of $\sqrt{r_3}$. } 
\label{fig:LLMrings} 
\end{center}
\end{figure}

With these basics set up, lets now recall the string theory result\cite{Hofman:2006xt}.
For well separated magnons, each magnon transforms in a definite $SU(2|2)^2$ representation. 
The closed string transforms as the tensor product of the individual magnon representations. 
To specify the representation of a magnon, we specify the central charges appearing in the $SU(2|2)^2$ algebra.
Think of the LLM plane as a complex plane.
Each magnon corresponds to a directed line segment on this plane.
Every directed line segment is equivalent to a complex number, $k$ with phase given by the direction of the line segment
and magnitude given by the length of the line segment.
The magnitude of $k$ is the length of the line corresponding to the magnon. 
The energy of the magnon is determined by supersymmetry to be
\bea
E =\sqrt{1+2g^2 |k|^2}=1+g^2 |k|^2+...\label{stres}
\eea
For a magnon which subtends an angle $\theta$ we have $|k|=4 R^2\sin^2{\theta\over 2}$ with $R$ the radius at
which the closed string is localized.
The angle $\theta$ is identified with the momentum $p$ of the magnon\cite{Hofman:2006xt}.
As we have just discussed, $R^2={N_{\rm eff}\over N}$ so that there is a perfect agreement between the gauge theory
result (\ref{gtres}) and the order $g^2$ contribution to the string theory result (\ref{stres}):
our proposal for operators dual to localized closed string states is correct.

\section{Another example and another method}\label{bigannulus}

The above results suggest that the only effect of working on an LLM background is to replace  
$g_{YM}^2 N\to g_{YM}^2 N_{\rm eff}$.
As we have just seen, this replacement in the gauge theory reproduces the central charge predicted by the string theory analysis.
In this section we want to explore what is perhaps the simplest setting in which this conclusion can be probed - simple
enough that we can do it without any restricted Schur polynomial technology.
This problem was considered in \cite{deMelloKoch:2009zm} and this section is just a quick review of those results.  
We will confirm our $g_{YM}^2 N\to g_{YM}^2 N_{\rm eff}$ conclusion for this simple example.

Consider a background $\chi_B(Z)$ with $B$ a Young diagram that has $N$ rows and $M$ columns, with $M$ of order $N$. 
We write
\bea
 \chi_B(Z) =\left(\det (Z)\right)^M\, .
\eea
The corresponding LLM boundary condition is a single black annulus hugging a central white disk of area $\pi M/N$.
The simplicity of this example is due to two facts:
\begin{itemize}
\item[1.] We have the well known formula for the derivative of a determinant
\bea
 {\partial\over\partial Z^i_j}\chi_B(Z)= M(Z^{-1})^j_i\chi_B(Z)\, .
\eea
\item[2.] In this background, multiplying by a trace is a local operation.
The background looks like a single annulus, so there is only one outer edge.
As a consequence, our local loops are
\bea
    O(\{ n_k\})) \chi_B(Z)
\eea
\end{itemize}

Thus, we can write
\bea
   D  \chi_B(Z) O(\{ n_k\})) = \chi_B(Z) D_{\rm eff} O(\{n_k\})
\eea
where
\bea
   D_{\rm eff} ={1\over\chi_B(Z)}D \chi_B(Z)=D- {Mg_{YM}^2\over 8\pi^2} 
\Tr \left( \left(ZYZ^{-1}+Z^{-1}YZ-2Y\right){\partial\over \partial Y}\right)
\eea
It is now simple to see that
\bea
&&D_{\rm eff}O(\{ n_k\})\cr
&=&{g_{YM}^2(N+M)\over 8\pi^2}
\left(2O(\{n_k\})-O(\{n_1+1,n_2-1,\cdots,n_m\})-O(\{n_1-1,n_2+1,\cdots,n_m\})
\right)\cr
&+&{g_{YM}^2(N+M)\over 8\pi^2}
\left(2O(\{n_k\})-O(\{n_1,n_2+1,n_3-1,\cdots,n_m\})-O(\{n_1,n_2-1,n_3+1,\cdots,n_m\})
\right)\cr
&&\quad \vdots
\cr
&+&{g_{YM}^2(N+M)\over 8\pi^2}
\left(2O(\{n_k\})-O(\{n_1-1,n_2,\cdots,n_m+1\})-O(\{n_1+1,n_2,\cdots,n_m-1\})
\right)\cr
&&
\eea
Since for this $B$ we have $N_{\rm eff}=N+M$, this is in perfect agreement with our expectations.

\section{Excitations on the inner edge of an annulus}\label{inneredge}

The loops we have constructed have been localized on the outer edge of a black disk or annulus on the LLM plane.
These excitations are created by adding boxes to inwardly pointing corners of the Young diagram describing the background.
It is also interesting to construct loops that are localized on the inner edge of a black annulus.
The inner edges correspond to outwardly pointing corners of the Young diagram describing the background.
It is not possible to add boxes at these corners; it is possible to remove boxes.
Thus, it is natural to describe these excitations by removing boxes.
Indeed, to orbit on the inner edge of a black ring these excitations must have negative the angular momentum carried
by a KK graviton dual to $Z$.
Thus these excitations must have an opposite ${\cal R}$ charge to that of $Z$; for each box we remove we do indeed decrease 
the ${\cal R}$ charge by 1.
One could also have considered including $Z^\dagger$s in the operator, which also decrease the ${\cal R}$ charge by 1.
However, including $Z^\dagger$s in the background also increases the dimension and takes us out of the class of
${1\over 2}$ BPS backgrounds.
``Empty box gravitons'' have been studied in \cite{Koch:2008ah} where they were called ``countergravitons''.
In Appendix \ref{emptyboxes} we have worked out enough details that we have indeed confirmed this expectation.
Using the resulting loop operators we again reproduce the string theory expectations.

\section{Conclusions}

In this article we have constructed operators dual to closed string states probing an LLM geometry.
We have not described the most general LLM geometry: we have focused on geometries arising when we color
the LLM plane with $O(1)$ concentric rings.
Further, we have computed the one loop anomalous dimensions of these operators.
Although our computations are quite technical the final result is remarkably simple: the action of the dilatation operator 
in the nontrivial geometry is given by simply scaling the $N$ dependence in the planar result.
We have only argued this at one loop, but we expect this to go through for higher loops.
Indeed, at $k$ loops, using the Casimir arguments developed in section 5, we know that at most $k$ boxes can shift position on 
the Young diagram and further that the background, labeled by Young diagram $B$, is not changed.
In this case, we again recover the planar dilatation operator but with $N$ scaled as it was at one loop level.
The intuition coming from the $SU(2|2)^2$ symmetry of the problem also supports this conclusion.
Indeed, from the one loop anomalous dimension we can read off the central charge of the magnon
and, thanks to supersymmetry, this completely determines the energy of the magnon.

\begin{figure}[ht]%
\begin{center}
\includegraphics[width=0.9\columnwidth]{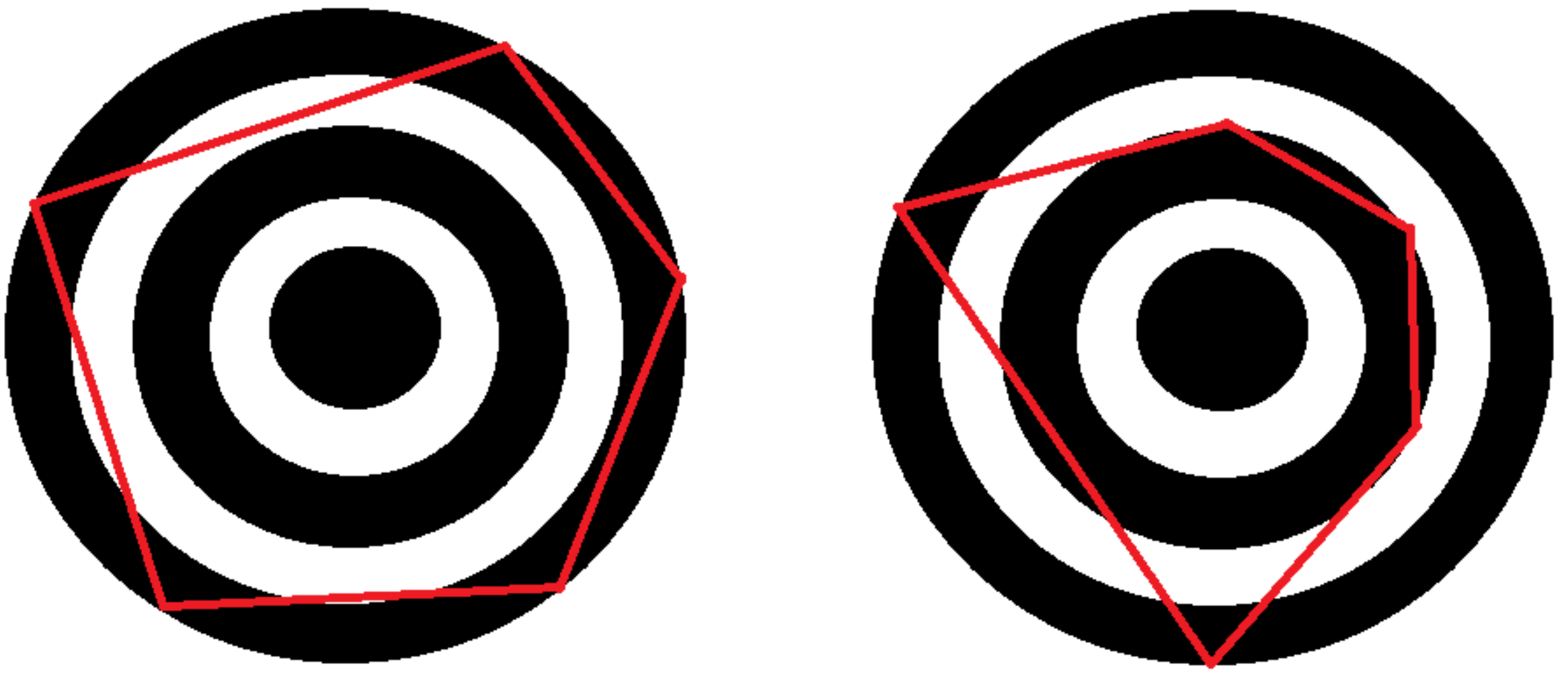}%
\caption{The string worldsheet on the left above corresponds to an integrable anomalous dimension problem in the dual gauge
theory. The problem on the right doesn't. Conservation of momentum implies the angle swept out by the magnons is the
same before and after scattering. Energy conservation is more difficult to describe geometrically. At very strong coupling 
it says that the sum of the lengths of the magnons scattering is the same before and after scattering.
It is clear that scattering of magnons for the worldsheet on the left is elastic. For the worldsheet on the right, when a magnon 
with endpoints on edges of different rings scatters, the scattering is in general inelastic.}%
\label{fig:rings}%
\end{center}
\end{figure}

For a closed string state described by a polygon with all vertices on the outer edge of a single ring, magnon scattering
is again elastic and the problem is integrable.
The spectrum of anomalous dimensions and even the expressions for the exact eigenstates can all be obtained from the
answers in the planar limit by simply replacing $N$ with $N_{\rm eff}$.
There are also closed string states with vertices located on the outer edges of distinct rings.
For these the magnon scattering problem is not elastic and the problem is not integrable.
However, thanks to the $SU(2|2)^2$ symmetry, the two body S-matrix can still be determined exactly, up to a phase.
It would be interesting to construct these S-matrices and verify their dynamical content.
For example, the poles of these S-matrices should contain information about the boundstate spectrum of the 
theory\cite{Dorey:2006dq}.

Previous studies of restricted Schur polynomials have focused on operators dual to states in the trivial AdS$_5\times$S$^5$
background.
In this article we have carried out a detailed study involving operators dual to states in a non-trivial geometry.
What lessons have we learned from this study?
Our construction of localized operators has involved a new ingredient which has not featured in any previous
constructions: the $Z$ fields belonging to the background have not been treated on the same footing as
the $Z$ fields belonging to the closed string excitation.
In the $SU(2)$ sector all previous constructions of restricted Schur polynomials have made use of the subgroup
$S_n\times S_m\subset S_{n+m}$.
The subgroup $S_n$ permutes indices belonging to $Z$s while $S_m$ permutes indices belonging to $Y$s.
The construction developed in this paper makes use of the subgroup $S_{n_B}\times S_{n_C}\times S_m\subset S_{n+m}$
when constructing operators in the $SU(2)$ sector.
The subgroup $S_{n_B}$ permutes indices belonging to the $Z$s making up the background, $S_{n_C}$ permutes indices
of $Z$s belonging to the closed string and $S_m$ permutes indices belonging to $Y$s.
Thus, the distinction between what is the background and what is the excitation is accomplished in the choice of the subgroup
and its representations.
This provides a concrete proposal for the gauge theory mechanism to distinguish between ``background'' and ``fluctuation''.
As a test of this proposal, we can try to determine where this construction breaks down and see if it matches our intuition.
The geometries we have studied arise from a boundary condition that consists of $O(1)$ fat black rings.
This is a geometry that is smooth on the string scale and we expect back reaction by stringy probes is negligible.
This is consistent with the fact that changes in the background Young diagram $B$ are suppressed in the large $N$ limit.
We could also consider a background $B$ described by a Young diagram with $O(N)$ corners.
The boundary condition for this geometry is a set of $N$ very thin rings, giving rise to a geometry that has
nontrivial features, even on the string scale.
We expect that this detailed and intricate geometry is disturbed by a stringy probe.
In this case there are so many outward pointing corners, that not all of them can be well separated from the 
local excitation.
In this case, the dilatation operator will start to mix operators with distinct background Young diagrams $B$, even at large $N$.
Consequently, taking $N\to\infty$ will no longer suppress backreaction, and we see that our gauge theory mechanism to distinguish
between ``background'' and ``fluctuation'' has failed precisely where we would expect it to.

Although small deformations of the ${1\over 2}$-BPS sector are special, this sector of the theory may
provide insight into more generic situations\cite{Balasubramanian:2005mg}.
For example, the presence of a horizon in the geometry signifies a region of high entropy - proportional to the horizon area.
To obtain a large entropy we need configurations that support many almost equivalent but different excitations, that could
provide the microstates responsible for this entropy.
It is natural to associate this with a region on the Young diagram that has many corners, since the number of possible excitations
is related to the number of representations of the subgroup we can subduce, and this is related to the number of 
boxes we can remove.
Indeed, the typical state in the ${1\over 2}$-BPS sector has $O(N)$ corners\cite{Balasubramanian:2005mg}.
For an excitation finding itself in this ``corner rich'' region, there are many outward pointing corners nearby the local excitation, 
so that it will spread over this region as time evolves.
This has the potential to provide a natural field theory origin for tidal forces and even the longitudinal string spreading 
effects experienced at the horizon (see for example \cite{Dodelson:2015toa}).
With some thought we expect that simple computations, to provide a quantitative interrogation for this identification, are 
possible.

It would seem to be straight forward to consider giant graviton excitations of these LLM geometries.
It appears that the arguments that work for the closed string excitations will go through, without modification, for the giants.
It would be interesting to check the details and verify that this is indeed the case.

Acting with $\chi_B(Z)$ on the vacuum produces the state that corresponds to the LLM background spacetime.
The loop $O(\{n_k\})$ operator creates a single closed string state in the AdS$_5\times$S$^5$ spacetime; it may have been
natural to expect that acting with $O(\{n_k\})$ on the state corresponding to the LLM geometry would produce a string
probing the LLM spacetime.
This is not the case: as we have seen, the product $O(\{n_k\})\chi_B(Z)$ does not produce a closed string localized
on the background spacetime.
However, taking a product of the representation describing the background and the representation describing the excitation
does indeed give the representation relevant for describing a closed string localized on the LLM spacetime.
The resulting operator involves a highly nontrivial mixing of the indices of the $Z$ fields belonging to the background and
$Z$ fields belonging to the excitation in the operator.
The simplest description of this mixing is in terms of the representations involved, so that the representation theory
description of the problem furnishes a very natural language.
We have also seen that the action of the dilatation operator is written in terms of the value of the factor at the
location on the Young diagram $R_B$ where the excitation is added.
This factor, which is a representation theoretic quantity, is related to the radius squared at which the excitation orbits on the 
LLM plane, a geometric quantity. 
In studies of open strings attached to giant gravitons 
\cite{Balasubramanian:2004nb,de Mello Koch:2007uu,de Mello Koch:2007uv,Bekker:2007ea}, each giant graviton
is described by a long row or column and each open string is associated with a box on the Young diagram.
The strength of the open string end point interactions is set by the factor of this box\cite{de Mello Koch:2007uv,Bekker:2007ea}.
Comparisons with a magnon description again gives exact agreement, with the factors combining to exactly
reproduce the magnon energy in the dual string theory description \cite{Koch:2015pga}.
It is striking how natural the language constructed using representation theory and Young diagrams is.

{\vskip 0.75cm}

\noindent
{\it Acknowledgements:}
We would like to thank Vishnu Jejjala and Sanjaye Ramgoolam 
for useful discussions and/or correspondence.
This work is based upon research supported by the South African Research Chairs
Initiative of the Department of Science and Technology and National Research Foundation.
Any opinion, findings and conclusions or recommendations expressed in this material
are those of the authors and therefore the NRF and DST do not accept any liability
with regard thereto.

\begin{appendix}

\section{Local Restricted Schur Polynomials}\label{localrschur}

The restricted Schur polynomials provide a complete basis for the local operators of the gauge theory.
If we restrict to the $SU(2)$ sector of the theory, these polynomials are labeled by three Young diagrams
and two multiplicity labels $\chi_{R,(r,s)\mu_1\mu_2}(Z,Y)$.
In this study our goal has been to understand how we can describe perturbations around a nontrivial LLM spacetime geometry.
Our results imply that one can write down ``local restricted Schur polynomials'' which provide a basis for such excitations.
The background is described by a Young diagram $B$ with order $N^2$ boxes.
In the $SU(2)$ sector of the theory, the perturbation can again be labeled by three Young diagrams and two multiplicity labels
$R,(r,s)\mu_1\mu_2$.
An economical way to summarize the construction of section \ref{LocalizedString} is to notice that the localized restricted Schur
polynomial uses representations\footnote{Recall that there are $n_B$ $Z$ fields making up the background geometry and $n_C+m$
fields in the loop.} of $S_{n_B+n_C+m}$ and of the subgroup $S_{n_B}\times S_{n_C}\times S_m$.
In terms of this language, we can write the localized restricted Schur polynomial as
\bea
O_{\hat{T}_B,(\hat{t}_B,u)\beta\alpha}(Z,Y)=O_{T_B,(B,T,t,u)\beta\alpha}(Z,Y)\label{genrsp}
\eea
In the above formula the triple of Young diagrams $B,t,u$ label an irrep of $S_{n_B}\times S_{n_C}\times S_m$. 
The labels $\beta\alpha$ as well as $T$ are multiplicity labels.
We have seen that taking a product of the operator describing the closed string and the operator describing the background 
geometry does not produce the desired closed string on the background spacetime.
In the operators (\ref{genrsp}) there is a simple tensor product between the representation (of $S_{n_B}$) describing the background the representation (of $S_m\times S_{n_C}$) describing the excitation.
This simple tensor product generates highly nontrivial mixing of the indices of the $Z$ fields belonging to the background and
$Z$ fields belonging to the excitation in the operator (\ref{genrsp}).

A number of identities valid for the restricted Schur polynomials also hold for the local restricted Schur polynomials.
The most important identity of this type is the completeness of local restricted characters in a specific background $B$.
The usual identity for the completeness of restricted characters, obtained in \cite{Bhattacharyya:2008xy}, reads
($r\vdash n$ and $s\vdash m$)
\bea
\sum_{R,(r,s)\alpha\beta}
{d_R\over d_r d_s (n+m)!}
\chi_{R,(r,s)\alpha\beta}(\tau)\chi_{R,(r,s)\beta\alpha}(\sigma)
=\delta_{[\sigma][\tau]}
\eea
The delta function on the right hand side is 1 if  we can satisfy
\bea
  \rho\tau\rho^{-1}=\sigma
\eea
for some $\rho\in S_n\times S_m$.
The completeness of the local restricted characters reads
\bea
\sum_{R,(r,s)\alpha\beta}
{d_R\over d_r d_s (n+m)!}
\chi_{\hat R_B,(\hat r_B,s)\alpha\beta}(\tau)\chi_{\hat R_B,(\hat r_B,s)\beta\alpha}(\sigma)
=d_B^2 \delta_{[\sigma][\tau]}\label{newid}
\eea
where $\sigma,\tau\in S_{n_C+m}$ and the delta function on the right hand side is 1 if  we can satisfy
\bea
  \rho\tau\rho^{-1}=\sigma
\eea
for some $\rho\in S_{n_C}\times S_m$.
Notice that in (\ref{newid}) the permutations $\sigma,\tau$ act only on the boxes belonging to the excitation
and the orthogonality in the permutations is achieved even though we only sum over the excitation labels.
It is completeness within a fixed background $B$.

\section{Restricted Character Identities}\label{identities}

In this section we would like to derive the exact identity for restricted characters, that reduces to
(\ref{restrictedCharId}) in the planar limit.
It is simple to verify that
\bea
  -{g_{YM}^2\over 8\pi^2}{\rm Tr}\left(
\left[ Y,Z\right]\left[{d\over dY},{d\over dZ}\right]\right)
{\rm Tr}(\sigma Y^{\otimes m} Z^{\otimes n})\cr
= {g_{YM}^2\over 8\pi^2}\sum_{p=1}^m\sum_{q=m+1}^{m+n}\Big(
\delta^{i_q}_{i_{\sigma (p)}}[Y,Z]^{i_p}_{i_{\sigma (q)}}
-\delta^{i_p}_{i_{\sigma (q)}}[Y,Z]^{i_q}_{i_{\sigma (p)}}\Big)
Y^{i_1}_{i_{\sigma(1)}}\cdots Y^{i_{p-1}}_{i_{\sigma(p-1)}} Y^{i_{p+1}}_{i_{\sigma(p+1)}}\cdots
 Y^{i_{m}}_{i_{\sigma(m)}}\cr
\times
Z^{i_{m+1}}_{i_{\sigma(m+1)}}\cdots Z^{i_{q-1}}_{i_{\sigma(q-1)}} Z^{i_{q+1}}_{i_{\sigma(q+1)}}\cdots
 Z^{i_{m+n}}_{i_{\sigma(m+n)}}\label{startexact}
\eea
The above formula is exact - it includes much more than just the planar contractions.
It is correct regardless of the way we scale $m$ and $n$ with $N$.
Further, the operator ${\rm Tr}(\sigma Y^{\otimes m} Z^{\otimes n})$ can have any trace structure - it is not
in general a single trace operator.
To obtain the large $N$ planar approximation, recall that all the $N$ dependence in this case comes from
contracting index loops and that the leading contribution comes from terms for which the dilatation operator
contracts a pair of indices to produce a factor of $N$.
Looking at the above expression, the terms that contribute in the planar approximation have $\sigma(p)=q$ or
$\sigma(q)=p$.
Keeping only these terms, the above expression can be written as
\bea
  -{g_{YM}^2\over 8\pi^2}{\rm Tr}\left(
\left[ Y,Z\right]\left[{d\over dY},{d\over dZ}\right]\right)
{\rm Tr}(\sigma Y^{\otimes m} Z^{\otimes n})
= {N g_{YM}^2\over 8\pi^2}\sum_{p=1}^m\sum_{q=m+1}^{m+n} (\delta_{q,\sigma (p)}+
\delta_{p,\sigma (q)})\cr \times\Big(
{\rm Tr}(\sigma Y^{\otimes m} Z^{\otimes n})
-{\rm Tr}\left((p,q)\sigma (p,q)Y^{\otimes m} Z^{\otimes n}\right)\Big)
\eea
Now, using the identity
\bea
{\rm Tr}(\sigma Y^{\otimes m} Z^{\otimes n})=\sum_{T,(t,u)\alpha\beta}
\sqrt{f_T{\rm hooks}_t{\rm hooks}_u\over {\rm hooks}_T}\chi_{T,(t,u)\alpha\beta}(\sigma^{-1})
O_{T,(t,u)\beta\alpha}(Z,Y)
\eea
and the exact action of the one loop dilatation operator, we find
\bea
&&\sum_{T,(t,u)\nu_1\nu_2}
\sqrt{f_T{\rm hooks}_u\over {\rm hooks}_{T/t}}\chi_{T,(t,u)\nu_2\nu_1}(\sigma^{-1})
N_{T,(t,u)\nu_1 \nu_2;R,(r,s)\mu_1\mu_2}={g_{YM}^2 N\over 8\pi^2}
\sqrt{f_R{\rm hooks}_s\over {\rm hooks}_{R/r}}\cr
&&\quad\times \sum_{p=1}^m\sum_{q=m+1}^{m+n}
\left(\delta_{q,\sigma (p)}+\delta_{p,\sigma (q)}\right)
\Big(
\chi_{R,(r,s)\mu_2\mu_1}(\sigma^{-1})-
\chi_{R,(r,s)\mu_2\mu_1}\big((p,q)\sigma^{-1}(p,q)\big)\Big)\cr
&&\label{EaxtRestrictedCharId}
\eea
This is in complete agreement with the identity (\ref{restrictedCharId}) between the restricted characters.

To go beyond the planar limit, return to (\ref{startexact}). After a little work, we obtain
\bea
  -{g_{YM}^2\over 8\pi^2}{\rm Tr}\left(
\left[ Y,Z\right]\left[{d\over dY},{d\over dZ}\right]\right)
{\rm Tr}(\sigma Y^{\otimes m} Z^{\otimes n})
=-{g_{YM}^2\over 8\pi^2}\sum_{p=1}^m\sum_{q=m+1}^{m+n}
{\rm Tr}\left( \psi {\bf 1}\otimes Y^{\otimes m}\otimes Z^{\otimes n}\right)\cr
\label{permres}
\eea
where $\psi$ depends on $p$ and $q$.
$\psi$ belongs to the group algebra of $S_{n+m+1}$, i.e. it is a linear combination of $S_{n+m+1}$ group elements.
We will use the notation $\psi^{-1}$ to denote the element of the group algebra obtained by summing the inverse of
each term summed to form $\psi$.
Concretely if $p\ne 1$ and $q\ne m+1$
\bea
\psi =(p,1)(q,m+1)\sigma (q,1,p,m+n+1,m+1)\cr
 - (1,q,m+1,p)\sigma (q,m+n+1,m+1)(p,1)\cr
-(1,p)(m+1,m+n+1,q)\sigma (p,m+1,q,1)\cr
+ (p,1,q,m+1,m+n+1)\sigma (m+1,q)(p,1)
\eea
If $p=1$ but $q\ne m+1$ we have
\bea
\psi=(q,m+1)\sigma (q,1,m+n+1,m+1)\cr
 - (1,q,m+1)\sigma (q,m+n+1,m+1)\cr
-(q,m+1,m+n+1)\sigma (1,m+1,q)\cr
+ (q,m+1,m+n+1,1)\sigma (m+1,q)
\eea
If $q=m+1$ but $p\ne 1$ we have
\bea
\psi =(p,1)\sigma (p,m+n+1,m+1,1)\cr
 - (p,1,m+1)\sigma (m+n+1,m+1)(p,1)\cr
-(1,p)(m+1,m+n+1)\sigma (p,m+1,1)\cr
+ (p,1,m+1,m+n+1)\sigma (p,1)
\eea
Finally, if $p=1$ and $q=m+1$ we have
\bea
\psi =\sigma (1,m+n+1,m+1)
 - (1,m+1)\sigma (m+n+1,m+1)\cr
-(m+1,m+n+1)\sigma (1,m+1)
+ (1,m+1,m+n+1)\sigma 
\eea

Now, on the right hand side of (\ref{permres}), replace ${\bf 1}$ by a new field $W$.
We will later turn $W$ back into ${\bf 1}$ by acting with ${\rm Tr}{\partial\over\partial W}$.
With $W$ on the right hand side of the above expression, we can write everything in terms of 
restricted Schur polynomials built using 3 complex fields, $W$, $Z$ and $Y$. 
These are labeled by 4 Young diagrams: ${\tiny\yng(1)}$ for $W$, $s\vdash m$ for the $Y$ fields, $r\vdash n$ for the $Z$ fields and 
$R^+\vdash {m+n+1}$ for the respresentation which mixes all three fields.
When removing a single box there is no multiplicity label.
Thus, we need a multiplicity for $s$ only.
In this way we obtain the following result
\bea
  -{g_{YM}^2\over 8\pi^2}{\rm Tr}\left(
\left[ Y,Z\right]\left[{d\over dY},{d\over dZ}\right]\right)
{\rm Tr}(\sigma Y^{\otimes m} Z^{\otimes n})\cr
=-{g_{YM}^2\over 8\pi^2}\sum_{p=1}^m\sum_{q=m+1}^{m+n}\sum_{R^+,r,s,\mu_1,\mu_2}
{d_{R^+} n! m!\over d_r d_s (n+m+1)!}
\chi_{R^+,(r,s,{\tiny \yng(1)})\mu_1\mu_2}\Big(\psi^{-1}\Big)
D_W \chi_{R^+,(r,s,{\tiny \yng(1)})\mu_2\mu_1}(Z,Y,W)\cr
=-{g_{YM}^2\over 8\pi^2}\sum_{p=1}^m\sum_{q=m+1}^{m+n}\sum_{R^+,r,s,\mu_1,\mu_2}
\sum_R {d_{R^+} c_{R^+ R} n! m!\over d_r d_s (n+m+1)!}
\chi_{R^+,(r,s,{\tiny \yng(1)})\mu_1\mu_2}\Big(\psi^{-1}\Big)
\chi_{R,(r,s)\mu_2\mu_1}(Z,Y)\cr
\label{exres}
\eea
Converting to normalized restricted Schur polynomials the above result gives the following identity for restricted characters
\bea
&&\sum_{T,(t,u)\nu_1\nu_2}
\sqrt{f_T{\rm hooks}_u\over {\rm hooks}_{T/t}}\chi_{T,(t,u)\nu_2\nu_1}(\sigma^{-1})
N_{T,(t,u)\nu_1 \nu_2;R,(r,s)\mu_1\mu_2}\cr
&&=-{g_{YM}^2\over 8\pi^2}\sum_{R^+}c_{R^+ R}
{{\rm hooks}_R\over {\rm hooks}_{R^+}}
\sqrt{f_R{\rm hooks}_s\over {\rm hooks}_{R/r}}
\sum_{p=1}^m\sum_{q=m+1}^{m+n}
\chi_{R^+,(r,s,{\tiny \yng(1)})\mu_2\mu_1}\Big(
\psi^{-1}\Big)\cr
&&\label{ExactRestrictedCharId}
\eea
This is an exact statement and no simplification for large $N$ has been used.

Localization of the boxes belonging to the closed string is accomplished because there is a permutation containing cycles 
which mix the $Y$ fields with the $Z$ fields. 
As we explained in section \ref{LocalizedString}, this cycle ``ties'' the $Z$ and $Y$ boxes together.
Since the $Y$ boxes are concentrated in one location on the Young diagram, the $Z$ boxes will be concentrated
there too.
The original localized loop is defined using a permutation $\sigma$ that is an $m+n_C$ cycle; this implies that all
of the fields belonging to the local excitation will be concentrated at one location on the Young diagram.
We want to explore the cycle structure of $\psi^{-1}$ which will determine whether or not the boxes associated
to the closed string in (\ref{ExactRestrictedCharId}) are localized or not, i.e. if the dilatation operator acts on a localized
loop operator, does it produce another localized loop operator?

$\psi$ and $\psi^{-1}$ have the same cycle structure, so we can focus on $\psi$.
$\psi$ is a sum of four terms that are conjugate to (and hence have the same cycle structure as) the following
four permutations
\bea
   \sigma (p,q) (m+n+1,q)\qquad    (p,q)\sigma (m+n+1,q)\cr
   (m+n+1,q) \sigma (p,q) \qquad    (m+n+1,q) (p,q)\sigma 
\eea
The permutation $\sigma$ is an $m+n_C$ cycle.
Permutations $(p,q)\sigma$ and $\sigma (p,q)$ correspond to a product of a $k$ cycle with an $n_C+m+1-k$
cycle with $k=1,2,\cdots n_C+m$\footnote{The planar contribution comes from the term with $k=1$ of the form
$(1)(\cdots)$ with $(\cdots)$ an $n_C+m$ cycle.}. 
The value of $k$ depends on the value of $p$ and $q$.
The sum over $p$ and $q$ produces $O(N)$ terms.
There are only $O(1)$ terms of a specific $k$, so the contribution for any given fixed $k$ can safely be neglected in the
large $N$ limit.
The $k=1$ term is enhanced with a factor of $N$; this is the term that dominates the planar limit.
Multiplying by the two cycle $(m+n+1,q)$ does not induce any further splitting because $m+n+1$ does not
appear in $\sigma$ or $(p,q)$.
In section \ref{LLMmag} we argued that localization of the $Y$ boxes after the dilatation operator acts is guaranteed.
Further, localization of all but one $Z$ box is also guaranteed.
The way that this simple box can move to a distant location is if $k=1$ and the label of the distant box sits in the 1-cycle
in $(p,q)\sigma$ or $\sigma (p,q)$.
As we have just explained, this term can be neglected at large $N$ so that the dilatation operator only mixes localized loop
operators.

\subsection{Numerical Check}

The formula (\ref{exres}) is rather interesting as it is the exact one loop dilatation operator written in the trace basis.
Such a formula has not been written in the literature before and it will provide a useful starting point for other large $N$ but
not planar expansions.
Indeed, we have used it for precisely this purpose in the LLM backgrounds.
Given the potential usefulness of such a formula, we have checked it numerically, for $n=3$ $Z$s and $m=2$ $Y$s.
In the case $m=2$ we do not need multiplicity labels and this simplifies the computations significantly.

\noindent
{\bf Evaluating restricted characters:} 
By $S_5$ denote the symmetric group that permutes 1,2,3,4 and 5.
By $S_3$ denote the symmetric group that permutes 3,4 and 5.
By $S_2$ denote the symmetric group that permutes 1 and 2.
Introduce the projectors
\bea
   P_s ={d_s\over 2!}\sum_{\sigma\in S_2}\chi_s(\sigma)\Gamma_{R^+}(\sigma)\cr
   P_r ={d_r\over 3!}\sum_{\sigma\in S_3}\chi_r(\sigma)\Gamma_{R^+}(\sigma)\cr
   P_R ={d_R\over 5!}\sum_{\sigma\in S_5}\chi_R(\sigma)\Gamma_{R^+}(\sigma)
\eea
The restricted character can be written as
\bea
   \chi_{R^+,(r,s,{\tiny \yng(1)})}\Big(\psi\Big)={\rm Tr}_{R^+}(P_s P_r P_R \Gamma_{R^+}(\psi))
\eea
This can be written in terms of normal characters as
\bea
   \chi_{R^+,(r,s,{\tiny \yng(1)})}\Big(\psi\Big)=
{d_s d_r d_R\over n! m! (n+m)!}\sum_{\sigma_1\in S_2}\sum_{\sigma_2\in S_3}\sum_{\sigma_3\in S_5}
\chi_s(\sigma_1)\chi_r(\sigma_2)\chi_R(\sigma_3)\chi_{R^+}(\sigma_1\sigma_2\sigma_3\psi)
\eea
where $R$ is obtained by dropping the $W$ box from $R^+$.
The fact that the restricted characters can be written in terms of normal characters is a direct consequence of the
fact that for the problem we are considering, there is no need for multiplicity labels. 
The numerical results for the restricted characters were checked by confirming the known orthogonality relation obeyed
by restricted characters.

\noindent
{\bf Evaluating the restricted Schur polynomials $\chi_{R,(r,s)}(Z,Y)$;} The bulk of the work is in computing the
restricted characters $\chi_{R,(r,s)}(\sigma)$ for $\sigma\in S_5$.
We can again express this restricted character in terms of normal characters as
\bea
   \chi_{R,(r,s,{\tiny \yng(1)})}\Big(\psi\Big)=
{d_s d_r \over n! m!}\sum_{\sigma_1\in S_2}\sum_{\sigma_2\in S_3}
\chi_s(\sigma_1)\chi_r(\sigma_2)\chi_R(\sigma_1\sigma_2\psi)
\eea

\noindent
{\bf Evaluating the factor $c_{R^+ R}$:} This factor is equal to $N$ plus the content of the box.
The content of the box is easily coded numerically using a Jucys-Murphy element as follows\cite{de Mello Koch:2007uu}
\bea
   c_{R^+R}=N+{1\over d_r}\sum_{i=1}^{m+n}\chi_{R^+,(r,s,{\tiny \yng(1)})}\Big(\Gamma_{R^+}(i,m+n+1)\Big)
\eea

Finally, some patience shows that the labels $R^+,(r,s,{\tiny \yng(1)})$ run over a total of 52 values.
The formula (\ref{exres}) has been verified for all possible $\sigma$ and $n=3$ and $m=2$.

\section{Excitations with empty boxes}\label{emptyboxes}

As explained in section \ref{inneredge}, excitations about the inner edge of a black ring are described by removing
boxes from the background Young diagram $B$.
Our initial discussion is for the geometry of section \ref{bigannulus} so that there is a single outward pointing corner.
In this case we don't need to worry about ensuring our excitation is local.
$B$ has $N$ rows and $M$ columns and $M=O(N)$.
Since each box in the Young diagram corresponds to a $Z$ field, we can remove boxes by taking derivatives with
respect to $Z$.
The simplest example of this process is the reduction rule for Schur polynomials\cite{deMelloKoch:2004crq,de Mello Koch:2007uu},
given by
\bea
  {\rm Tr}\left({d\over dZ}\right)\chi_R(Z)=\sum_{R'}c_{RR'}\chi_R(Z)
\eea
with the sum over $R'$ running over all $R'$ that can be obtained by dropping a single box from $R$.
General formulas for the action of multitraces of derivatives are worked out in \cite{Koch:2008cm}.

\subsection{Schur Polynomials with empty boxes}

We would like to remove $n$ $Z$ boxes.
The removed boxes are then organized with a representation $r\vdash n$ or into a trace labeled by some
permutation $\sigma$.
The empty $Z$ boxes are generated by acting on some background described by Young diagram $B$.
A very natural guess for the operator which organizes the empty boxes into a single trace described by $\sigma$ is
\bea
   (\partial_Z)^{i_1}_{i_{\sigma(1)}}\cdots (\partial_Z)^{i_n}_{i_{\sigma(n)}}\chi_B(Z)=
{1\over (n_B-n)!}\sum_{\rho\in S_{n_B}}\chi_B(\rho\sigma^{-1}){\rm Tr}(\rho {\bf 1}^{\otimes n}Z^{\otimes n_B-n})
\eea
where
\bea
   (\partial_Z)^{i}_{j}={d\over dZ^{j}_{i}}
\eea
Replace the identity matrices in the above expression with a new field $W$ and turn these back into the identity by
acting with ${\rm Tr}\partial_W$.
Then use the completeness of the restricted Schur polynomials as well as the reduction rule for the restricted Schur polynomials
to obtain
\bea
(\partial_Z)^{i_1}_{i_{\sigma(1)}}\cdots (\partial_Z)^{i_n}_{i_{\sigma(n)}}\chi_B(Z)&=&
{1\over n!}({\rm Tr}\partial_W)^n
{1\over (n_B-n)!}\sum_{\rho\in S_{n_B}}\chi_B(\rho\sigma^{-1}){\rm Tr}(\rho \,\, W^{\otimes n}Z^{\otimes n_B-n})
\cr
&=&{1\over n!}({\rm Tr}\partial_W)^n
{1\over (n_B-n)!}\sum_{\rho\in S_{n_B}}\chi_B(\rho\sigma^{-1})\sum_{R,(r,s)\alpha\beta}
{d_R(n_B-n)!n!\over d_r d_s n_B!}\cr
&&\times \chi_{R,(r,s)\alpha\beta}(\rho^{-1})\chi_{R,(r,s)\beta\alpha}(Z,W)\cr
&=& \sum_{s\vdash n} \chi_s(\sigma) f_s(M)\chi_{B/s}(Z)\label{traceres}
\eea
To spell out the notation, we give an example of $B$, $s$ and $B/s$ below
\bea
B={\tiny \yng(6,6,6,6,6,6,6,6,6)}\qquad\qquad
s={\tiny \yng(4,1)}\qquad\qquad
B/s={\tiny \yng(6,6,6,6,6,5,5,5,4)}
\eea
The notation $f_r(M)$ stands for the product of factors of Young diagram $r$, but with $N$ replaced by $M$.
Thus, for example
\bea
   f_{\tiny\yng(3,2)}     = N^2 (N-1)(N+1)(N+2)\qquad\qquad
   f_{\tiny\yng(3,2)}(M)= M^2 (M-1)(M+1)(M+2)
\eea
Now consider the operator which organizes the empty boxes into some irrep $r\vdash n$. 
The operator which naturally achieves this is
\bea
\chi_{B;r}(Z,\partial_Z)={1\over n!}\sum_{\sigma\in S_n}
\chi_{r}(\sigma)
\partial_Z{}^{i_{1}}_{i_{\sigma (1)}}\cdots \partial_Z{}^{i_{n}}_{i_{\sigma (n)}}
\chi_B(Z)
\label{opdef}
\eea
Using (\ref{traceres}) and the orthogonality of characters, we have
\bea
\chi_{B;r}(Z,\partial_Z)=f_r(M)\chi_{B/r}(Z)
\eea
A few helpful formulas that follow from the results of this section are
\bea
\langle\chi_{B,r}(Z) \chi_{B,s}(Z)^\dagger\rangle &=& f_B f_r(M)\delta_{rs}\cr
\chi_{B,r}(Z)&=&{1\over n!}\sum_{\sigma\in S_n}\chi_r(\sigma ){\rm Tr}(\sigma \partial_Z^{\otimes n})\chi_B(Z)\cr
{\rm Tr}(\sigma \partial_Z^{\otimes n})\chi_B(Z)&=&\sum_{r\vdash n}\chi_r(\sigma)\chi_{B,r}(Z)
\eea

\subsection{Restricted Schur Polynomials with Empty Boxes}

In this section the operators we study belong to the $SU(2)$ sector and are built from $Z$ and $Y$ fields.
We would like to remove $Z$ boxes and organize them with a representation $r\vdash n$, add $Y$ boxes organized 
with a representation $s\vdash m$ and then organize the $Y$ and $Z$ boxes with a representation $R\vdash m+n$.
The operator which naturally achieves this is
\bea
\chi_{B;(B/R,r,s)\alpha\beta}(Z,Y,{\bf 1})=&&{1\over (n_B-n-m)!n!m!}\cr
&&\sum_{\sigma\in S_{n_B}}
\chi_{B;(B/R,r,s)\alpha\beta}(\sigma)
{\rm Tr}(\sigma  {\bf 1}^{\otimes n}Y^{\otimes m}Z^{\otimes n_B-n-m})
\label{opdef}
\eea
The empty $Z$ boxes are again generated by acting on the background described by Young diagram $B$.
Thus, four irreps play a role: $s\vdash m$ organizing the $Y$s, $r\vdash n$ organizing the empty boxes and $B/R$ organizing 
the $Z$ fields left in the background and $B$ the original background irrep.
To spell out how the different irreps are embedded in $B$, remove the empty boxes first and then remove the $Y$ boxes.
This implies that there are no multiplicities for either $r$ (thanks to fact that we are considering an outward pointing corner) or $B/R$
(always the case); the only multiplicity label needed is for $Y$ so the multiplicity labels take the same values as they did for 
the restricted Schur polynomial.
Thus, at infinite $N$ (where we don't need to worry about any possible cut off on the size of the Young diagram)
the number of restricted Schur polynomials is equal to the number of restricted Schur polynomials with empty boxes.
By introducing an extra field $W$, taking derivatives and using the reduction rule for restricted Schur polynomials,
we can write
\bea
\chi_{B;(B/R,r,s)\alpha\beta}(Z,Y,{\bf 1})&=&{1\over n!}\left({\rm Tr}\partial_W\right)^n
\chi_{B;(B/R,r,s)\alpha\beta}(Z,Y,W)\cr
&=&{d_r\over n!}f_r(M)\chi_{B/r,(B/R,s)\alpha\beta}(Z,Y)\label{etoftranslate}
\eea
It is now straight forward to compute the two point correlation function
\bea
T&=&\langle\chi_{B;(B/R,r,s)\alpha\beta}(Z,Y)\chi_{B;(B/T,t,u)\gamma\delta}(Z,Y)^\dagger \rangle\cr
&=&\delta_{RT}\delta_{rt}\delta_{su}\delta_{\beta\delta}\delta_{\alpha\gamma}
{f_B f_r(M)\over ({\rm hooks}_r)^2}{{\rm hooks}_{B/r}\over {\rm hooks}_{B/R}{\rm hooks}_u}
\eea

\subsection{Action of the Dilatation Operator}

It is rather simple to compute the action of the dilatation operator on restricted Schur polynomials with empty boxes.
Indeed, the label for the empty boxes is simply a short hand and we can easily translate these labels into the usual
description in terms of a representation for the $Z$ fields and one for the $Y$ fields.
This translation is given in (\ref{etoftranslate}).
This translation is nontrivial.
For example, one may have expected that the operator built using no empty slots and only $Y$s is BPS.
This is not the case.
A very simple example of this is (in this example no multiplicity labels are needed)
\bea
   D\chi_{B;(B/{\tiny\yng(2)},\cdot,{\tiny\yng(2)})}(Z,Y,{\bf 1})={g_{YM}^2 M\over 4\pi^2}
\chi_{B;(B/{\tiny\yng(2)},\cdot,{\tiny\yng(2)})}(Z,Y,{\bf 1})
-{g_{YM}^2 M\over 4\pi^2}\chi_{B;(B/{\tiny\yng(1,1)},\cdot,{\tiny\yng(1,1)})}(Z,Y,{\bf 1})\cr
\eea
Our main interest is in defining localized loop operators in this background.
To do this, it is more useful to use the following representation of the background
\bea
   \chi_B(Z)=(\det (Z))^M
\eea
as well as the derivative formula
\bea
   (\partial_Z)^i_j \det Z=(Z^{-1})^i_j\det Z
\eea
We see that each action of the derivative produces a factor of $Z^{-1}$.
In our previous formulas, $Z^{-1}$ did not participate.
As soon as we have more than $N$ fields, there are trace relations so that in this case there isn't a unique expression for
generic multitrace operators, so this apparent discrepancy should not worry us.
The natural structure for the loop operator is
\bea
O_B(\{ n_k\})&=& M^{-n}\Tr(\partial_Z^{n_1}Y\partial_Z^{n_2}Y\cdots \partial_Z^{n_m}Y)\chi_B(Z)\cr
                 &=&\Tr(Z^{-n_1}Y Z^{-n_2}Y\cdots Z^{-n_m}Y)\chi_B(Z)
\eea
where there are $m$ $Y$s and $n=n_1+n_2+\cdots+n_m$ $Z$s is the trace.
The second line above is only true in the large $N$ limit, where we can ignore splitting of the trace.
Rewrite the dilatation operator as (derivatives in $D$ do not act on fields inside $D$)
\bea
  D={g_{YM}^2\over 8\pi^2}\Tr \Big( \big[\left[Y,Z\right],\partial_Z\big]\partial_Y \Big)
\eea
When acting on $O_B(\{ n_k\})$, $D$ will produce $m$ terms obtained by replacing each $Y$ (a different one in each term)
in $O_B(\{ n_k\})$ with $\big[\left[Y,Z\right],\partial_Z\big]$.
After allowing all of the derivatives with respect to $Z$ to act on $\chi_B(Z)$ we obtain
(this result is not exact; it is obtained in the large $N$ limit with $n_C+m\sim O(\sqrt{N})$)
\bea
&&   DO_B(\{ n_k\})\cr
&&={g_{YM}^2 M\over 8\pi^2}
\left(2O_B(\{n_k\})-O_B(\{n_1+1,n_2-1,\cdots,n_m\})-O_B(\{n_1-1,n_2+1,\cdots,n_m\})
\right)\cr
&&+{g_{YM}^2 M\over 8\pi^2}
\left(2O_B(\{n_k\})-O_B(\{n_1,n_2+1,n_3-1,\cdots,n_m\})-O_B(\{n_1,n_2-1,n_3+1,\cdots,n_m\})
\right)\cr\cr
&&+ \qquad\qquad\cdots\cdots\cdots\qquad\qquad+
\cr\cr
&&+{g_{YM}^2 M\over 8\pi^2}
\left(2O_B(\{n_k\})-O_B(\{n_1-1,n_2,\cdots,n_m+1\})-O_B(\{n_1+1,n_2,\cdots,n_m-1\})
\right)\cr
&&\eea
which matches (\ref{actonloop}) perfectly and confirms our identification of the loop operator.
Further, we see that $N$ is replaced by $M$ which again reproduces the string theory expectations.

\subsection{More general backgrounds}

To construct localized restricted Schur Polynomials with empty boxes on the general multiring geometry, we need to
use localized derivative operators, that only remove boxes from a single corner of $B$.
Derivatives of this type have been defined and studied in \cite{Koch:2008ah}. 
Replacing $N$ by the factor of the outward pointing corner again reproduces the string theory expectations.

\end{appendix}

\end{document}